\documentclass[subeqn,epsf,psfig,aps,twocolumn,nofootinbib,superscriptaddress]{revtex4}

\usepackage{amsmath}
\usepackage{amssymb}

\newcommand{\Mpl}{M_{\rm{pl}}}

\def\spose#1{\hbox to 0pt{#1\hss}}
\def\simlt{\mathrel{\spose{\lower 3pt\hbox{$\mathchar"218$}}
     \raise 2.0pt\hbox{$\mathchar"13C$}}}
\def\simgt{\mathrel{\spose{\lower 3pt\hbox{$\mathchar"218$}}
     \raise 2.0pt\hbox{$\mathchar"13E$}}}
\def\simpropto{\mathrel{\spose{\lower 3pt\hbox{$\mathchar"218$}}
     \raise 2.0pt\hbox{$\propto$}}}

\usepackage{epsfig}
\usepackage{float,times}

\begin{document}
\title{Constraining $f(R)$ Gravity as a Scalar Tensor Theory}
\author{Thomas Faulkner}
\affiliation{Dept.~of Physics, Massachusetts Institute of Technology, Cambridge, MA 02139}
\author{Max Tegmark}
\affiliation{Dept.~of Physics, Massachusetts Institute of Technology, Cambridge, MA 02139}
\affiliation{MIT Kavli Institute for Astrophysics and Space Research, Cambridge, MA 02139}
\author{Emory F. Bunn}
\affiliation{Physics Department, University of Richmond, Richmond, VA 23173}
\affiliation{MIT Kavli Institute for Astrophysics and Space Research, Cambridge, MA 02139}
\author{Yi Mao}
\affiliation{Dept.~of Physics, Massachusetts Institute of Technology, Cambridge, MA 02139}
\begin{abstract}
We search for viable $f(R)$ theories of gravity, making use of the
equivalence between such theories and scalar-tensor gravity.
We find that models can be made consistent with solar system
constraints either by giving the scalar a high mass or by exploiting the
so-called chameleon effect.
However, in both cases, it appears likely that any late-time cosmic
acceleration will be observationally
indistinguishable from acceleration caused by a cosmological constant.
We also explore further observational constraints from, e.g., big
bang nucleosynthesis and inflation.
\end{abstract}
\maketitle

\section{Introduction}


Although the emerging cosmological standard model fits measurements 
spectacularly well (see \cite{Spergel:2006hy,Tegmark:2006az} 
for recent reviews),
it raises three pressing questions: what is the physics of the 
postulated dark matter, dark energy and inflation energy?
The need to postulate the existence of as many as three new substances 
to fit the data has 
caused unease among some cosmologists 
\cite{Peebles:1999eb,Peebles:2000ay,Sellwood:2000wk,Tegmark:2001zc}
and prompted concern that these complicated dark matter flavors constitute a
modern form of epicycles.
Our only knowledge about these purported substances comes from their gravitational effects.
There have therefore been numerous suggestions that
the apparent complications
can be eliminated by modifying the laws of 
gravity to remove the need for dark matter 
\cite{Milgrom:1983ca,Bekenstein:2004ne},
dark energy 
\cite{Boisseau:2000pr,Esposito-Farese:2000ij,Carroll:2003wy}
and inflation \cite{Starobinsky:1980te}, and perhaps 
even all three together \cite{Liddle:2006qz}.
Since attempts to explain away dark matter with modified gravity have been severely challenged by recent observations, 
notably of the so-called bullet cluster \cite{Clowe:2006eq}, 
we will focus on dark energy (hereafter ``DE'') and inflation.

There is also a second motivation for exploring alternative gravity theories: 
observational constraints on parametrized departures from general relativity (GR) have provided increasingly precise tests of 
GR and elevated confidence in its validity \cite{Will:1993ns,Will:2005va}.

\subsection{$f(R)$ gravity}

An extensively studied generalization of general relativity involves modifying 
the Einstein-Hilbert Lagrangian
in the simplest possible way, replacing $R - 2\Lambda$ by a more general 
function $f(R)$
\footnote{For the general case where $f$ depends on the full Riemann tensor $R^{\mu}_{\nu\alpha\beta}$
rather than merely on its contraction into the Ricci scalar $R$, 
this program is more complicated; a subset of these 
theories which are ghost
free can be written as $f(R,G)$, where $G = 
R^\beta_{\hphantom{\beta}\mu\nu\alpha} R_\beta^{\hphantom{\beta}\mu\nu\alpha}
- 4 R^{\mu\nu} R_{\mu\nu} + R^2$ is the Gauss-Bonnet scalar in 
4 dimensions \cite{Navarro:2005da}.
These theories lack a simple description
in terms of canonical fields; there is no so-called Einstein Frame. 
Progress has nevertheless been made along these lines,
and such Lagrangians may have more relevance to DE \cite{Navarro:2005da,Mena:2005ta,Navarro:2005gh} than ones independent of $G$.} 
\cite{Carroll:2003wy,Dick:2003dw,Dolgov:2003px,
Chiba:2003ir,Amendola:2006kh,
Capozziello:2006dj,Nojiri:2006gh,Olmo:2005hc,
Clifton:2005aj,Song:2006ej,Bean:2006up,
Erickcek:2006vf, Navarro:2006mw, Chiba:2006jp}.
The equations of motion derived from this Lagrangian 
differ from Einstein's field equations when $f(R)$ is nonlinear, but
the theory retains the elegant property of general coordinate invariance. 
In such a theory, the acceleration of our universe may be explained 
if $f(R)$ departs from linearity at small $R$, corresponding to 
late times in cosmological evolution.
In this case it may be possible to avoid invoking a cosmological
constant to explain cosmic acceleration, although one then
replaces the problem of a small cosmological constant with the
problem of no cosmological constant. 
In such models, the effective DE is dynamic, i.e., it is not
equivalent to a cosmological constant, leading to potentially interesting
observational signatures.
We refer
to these models as $f(R)$-DE theories. 

In addition to potentially explaining late-time acceleration, $f(R)$
theories may be relevant to early-universe physics, particularly
if $f(R)$ is non-linear at large $R$ 
\cite{Starobinsky:1980te,Schmidt:5}.
More generally,
it is of interest to study
$f(R)$ theories
because they are arguable the simplest setting in which one can attack
the
general question of which modified theories of gravity are allowed.

\subsection{The equivalence with scalar tensor gravity}


The modified Einstein field equations (and so the new Friedmann equation)
resulting from a non-linear $f(R)$ in the action can be seen
simply as the addition of a new scalar degree of freedom. 
In particular, it is well-known that 
these theories are exactly
equivalent to a scalar-tensor theory
\cite{Whitt:1984pd,
Jakubiec:1988ef}. It is therefore no surprise that 
for $f(R)$-DE theories, 
it is this scalar 
which drives the DE.
Before reviewing the mathematics of this equivalence in full detail in section \ref{sec:equiv},
we will discuss some important qualitative features below.

One can discuss the theory
in terms of the original metric $g_{\mu\nu}$,
in which case the degrees of freedom are not manifest.  Alternatively, 
by a 
conformal relabeling, one can reveal the theory to be regular gravity
$\tilde{g}_{\mu\nu}$, plus scalar field $\phi$. The former viewpoint is
referred to as the Jordan Frame (JF) and the latter as the Einstein Frame (EF).
Here
$\phi$ has the peculiar feature that in the JF, it exactly determines the
Ricci scalar $R$ and vice versa. So in the JF, the
Ricci scalar can in a sense be considered a non-canonical yet dynamical scalar field.
This feature is absent in normal general relativity,
where $R= - T/\Mpl^2 + 4 \Lambda$
is algebraically fixed by the trace of the stress energy tensor $T$.
Working in either frame
is satisfactory as long as one is careful about
what quantities are actually measurable, but we will find 
that the EF is much more useful for most of our calculations.

The coupling of the scalar field to matter is fixed in
$f(R)$ gravity, 
and is essentially of the same strength as the coupling of the graviton to
matter, except for the important case of massless conformally
invariant fields, which do not couple to $\phi$ at all.  
The dynamics of the theory are completely specified by the 
potential
$V(\phi)$ for
the scalar field in the EF, which is uniquely
determined by the functional form of $f(R)$.

\subsection{The $R-\mu^2/R$ example}


Such a scalar field is not without observational consequence for
solar system tests of gravity, especially for $f(R)$-DE models.
For \emph{any} scalar field driving DE, we can come to the following
conclusions:
First, the field value $\phi$ must vary on a time scale of order 
the Hubble time $H_0^{-1}$, if the DE is distinguishable
from a cosmological constant
(for longer time scale,  
the DE looks like a cosmological constant; for shorter time scales, 
we no longer get acceleration). 
On general grounds, such a scalar field must have a mass
of order $m_\phi^2 \sim H_0^2$. Second, the Compton wavelength of this scalar
field is on the order of the Hubble distance, so it will
mediate an attractive fifth force which is distinguishable from gravity
by the absence of any coupling to light.
Unless the coupling to matter is tiny compared to that
of gravity, many solar system based tests
of gravity would fail, such as measurements of the bending of light 
around the Sun \cite{Will:1993ns,Will:2005va}.

The archetypal example of $f(R)$-DE suffers from problems
such as these.
This model invokes the function 
\cite{Carroll:2001bv},
\begin{equation}
\label{eq:arch}
f(R) = R - \frac{\mu^4}{R}
\end{equation}
for $\mu \approx H_0$. This 
gives a $V(\phi)$ in the EF with a runaway exponential potential
at late times: 
$V(\phi) \sim H_0^2 \Mpl^2 \exp(-(3/\sqrt{6})\phi/\Mpl)$ (here large $\phi$ 
means small $R$ which means late times.)
With no matter, 
such a potential in the JF 
gives rise to an accelerating
universe with the equation of state parameter $w_X = -2/3$
\cite{Carroll:2001bv}.
This model, however, is riddled with problems.
First, the theory does not pass solar system tests 
\cite{Chiba:2003ir,Dick:2003dw,Dolgov:2003px,Erickcek:2006vf},
and second, the cosmology is inconsistent with observation
when non-relativistic matter is present 
\cite{Amendola:2006kh}. Both problems can be understood
in the dual scalar tensor theory. 

For cosmology, during
the matter dominated phase but at high redshifts,
the influence on the dynamics of $\phi$ from
the potential $V$ is small compared to the influence
from the coupling 
to matter, which manifests itself in terms of an effective potential
for $\phi$ of the form
\begin{equation}
V_{\rm{eff}}(\phi) = 
V(\phi) + \bar{\rho}_{\rm{NR}} \exp\left( - \frac{
\phi}{ \sqrt{6} \Mpl} \right),
\end{equation}
where $\bar{\rho}_{NR}$ is the energy density of non-relativistic (NR) matter.
(More
details of the exact form of this potential will be presented in 
the next section.) The second term dominates because $H_0^2 \Mpl^2 \ll
\rho_{\rm{NR}}$ and 
$\phi$ then rolls down the potential generated by $\rho_{\rm{NR}}$ and not
$V$. The result is that the universe is driven away from
the expected matter dominated era (MDE) into a
radiation dominated expansion in the JF with $H^2 \propto a^{-4}$,
after which it crosses directly into the accelerating phase, with
expansion driven by DE with an effective equation of state parameter $w=-2/3$.
This special radiation-dominated-like 
phase (which is not driven by radiation) was
dubbed the $\phi$MDE by \cite{Amendola:2006kh}, where it
was made clear that 
this phase is inconsistent with observation. We say that this
potential $V$ is unstable to large cosmological non-relativistic densities.

For the solar system tests, the potential $V(\phi)$ is also negligible, so 
the theory behaves exactly like a scalar-tensor theory with
no potential. Because the coupling to matter has the same strength
as that to gravity, the scalar field mediates a long-range fifth force,
and the theory is ruled out by solar system tests.
In particular,
\cite{Chiba:2003ir} found that $\gamma=1/2$ in the PPN framework,
which is in gross violation of the experimental bound.

The above solar system tests also seem to rule
out more general classes of $f(R)$-DE models 
\cite{Olmo:2005hc,Chiba:2006jp,Erickcek:2006vf,Navarro:2005gh,Navarro:2006mw}.
However on the cosmology front, 
it seems that one \emph{can} cook up 
examples of $f(R)$ consistent with some dynamical dark energy
\cite{Capozziello:2006dj,Nojiri:2006gh,Song:2006ej}: by demanding that
the cosmological expansion $a(t)$ take a certain form, one can integrate
a differential equation for the function $f$ that by design 
gives a universe with any desired expansion history $a(t)$.
In this way, one gets around the cosmological instability of the 
archetypal model mentioned above.
However, these functions are arguably very contrived,
and further investigation of solar system predictions are required to determine 
whether these models are viable.

\subsection{What $f(R)$-theories are allowed?}


We now try to find viable $f(R)$ theories by examining what \emph{is}
acceptable on the scalar tensor
side. We focus
on theories that pass solar system tests.
Because 
the coupling of the scalar field 
to matter is fixed in $f(R)$ theories, and the only freedom we have is with the potential
$V$, we must choose $V$ in such a way as 
to hide the scalar field
from the solar system tests that caused problems
for the models described above. We are aware of only two ways to do this.
The first is the Chameleon scalar field, which
uses non-linear effects from a very specific singular 
form of potential
to hide the scalar field from current tests of gravity 
\cite{Khoury:2003aq,Khoury:2003rn}.
The second is simply to give the scalar field a quadratic potential
with mass $m_\phi \gtrsim 10^{-3} \rm{eV}$,
so that the fifth force has an extent less than $0.2\rm{mm}$ and so
cannot be currently measured by laboratory searches for a fifth
force \cite{Hoyle:2000cv}.

We will find simple $f(R)$ models which reproduce these two types
of potentials and so by design pass solar system tests.
Finding functions $f$ which give \emph{exactly} these potentials
will simply generate models which are
indistinguishable from their scalar-tensor equivalent. However if
we search for simple choices of $f$ that reproduce these
potentials in a certain limit, then these theories
will not be exactly equivalent and might have distinguishable
features.


The Chameleon type $f(R)$ model seems to be the most plausible
model for attacking DE, as at first glance it seems to get
around the general problems mentioned above.
Indeed, 
one Chameleon model will arise quite naturally from a simple choice
of $f$. However,
we will show that the solar system constraints on this model
preclude any possible interesting late-time cosmological
behavior: the acceleration is observationally indistinguishable from a cosmological
constant.
In particular, for all the relevant physical situations 
this Chameleon model is the same as has been
considered before with no distinguishing features. However,
this model might provide clues in a search for
viable $f(R)$ theories that pass solar system tests and that may
give interesting late-time behavior. 

In an independent recent analysis, \cite{Navarro:2006mw} also discussed
the Chameleon effect in $f(R)$ theories. They focus on a slightly 
different set of Chameleon potentials and come to similar
conclusions. Their results and ours together suggest
that the Chameleon effect may be generic
to $f(R)$ theories.


We now turn from attempts to explain DE in $f(R)$ models to an arguably more
plausible scenario, which is simply to give the scalar
field a large mass.  These models
have no 
relevance for dynamic DE, but they do
have interesting consequences for
early universe cosmology.
The most theoretically best motivated functions, namely polynomials in $R$,
fit this class of $f(R)$ theories. The aim of this investigation 
is to explore
what we can possibly know about the function $f$. Because
this
question is very general, we will restrict our attention to a sub-class
of plausible $f(R)$ models. 

For these polynomial models, we will investigate possible
inflationary scenarios where the scalar partner $\phi$ is
the inflaton. We find the relevant model parameters which seed the
fluctuations of the CMB in accordance with experiment. We then
investigate general constraints on the model parameters
where $\phi$ is not an inflaton. We use solar system tests, 
nucleosynthesis constraints
and finally
an instability which is present in these theories when another
slow roll inflaton $\psi$ is invoked to explain CMB fluctuations.
This instability is analogous to that of the $\phi$MDE 
described above.


The rest of this paper is organized as follows.
In section \ref{sec:equiv} we review the equivalence of $f(R)$ theories with
scalar tensor theories, elucidating all the essential points we
will need to proceed. Then in sections \ref{sec:cham} and \ref{sec:mass}
we explore the Chameleon model and massive theories, respectively, focusing on observational constraints.
We summarize our conclusions in section \ref{sec:concl}.

\section{$f(R)$ duality with Scalar Tensor theories \label{sec:equiv}}

We study the ``modified'' gravity theory defined by the action
\begin{equation}
\label{eq:theaction}
S_{JF} = \int d^4x \sqrt{-g} \frac{\Mpl^2}{2} f(R) +
S_M\left[g_{\mu \nu}, \Psi, A_\alpha, \ldots\right]
\end{equation}

Here we present a run
down of the 
map to the scalar tensor theory; displaying the most
important points needed to proceed.
See for example 
\cite{Jakubiec:1988ef,Hindawi:1995cu,Carroll:2003wy} for more details
of the equivalence with scalar tensor theories. 

We \emph{choose} to 
fix the connection in $R$ as the Christoffel symbols and not
an independent field, as opposed to the Palatini
formalism - which results in a very different theory
\cite{Vollick:2003aw,Flanagan:2003rb,
Flanagan:2003iw,Carroll:2006jn,Amarzguioui:2005zq}.  

If one simply varies the action Eq.~(\ref{eq:theaction}) with respect
to the metric $g_{\mu\nu}$, then a fourth order equation for the
metric results. One can argue (using general coordinate invariance)
that the degrees of freedom in the field $g_{\mu\nu}$ can be split
into a massless spin-2 field $\tilde{g}_{\mu\nu}$ and
a massive scalar field $\phi$ with second order equations of motion.
This split is easily revealed at the level of the action. 
Following for example \cite{Hindawi:1995cu} we introduce
a new auxiliary scalar field $Q$ (a Lagrange multiplier) the
gravity part of Eq.~(\ref{eq:theaction}) may be written as
\begin{equation}
\label{eq:gravaction1}
S_{\rm{grav}} = \int d^4x \sqrt{-g} \frac{\Mpl^2}{2}
\left( f'(Q) (R - Q) + f(Q) \right)
\end{equation}
As long as $f''(Q) \neq 0$, the equation of motion ($\delta / \delta Q$) gives
$Q=R$ and Eq.~(\ref{eq:gravaction1}) becomes the original gravity action.
This may be written in the more suggestive form 
\begin{equation}
S_{\rm{grav}} = \int d^4x \sqrt{-g} \left(
\frac{\Mpl^2}{2}
\chi R  - \chi^2 V(\chi) \right)
\end{equation}
by relabeling
$f'(Q) \equiv \chi$. 
This is a scalar tensor theory of gravity 
with Brans Dicke parameter $\omega_{BD}=0$ \cite{Brans:1961sx} 
and potential \cite{Hindawi:1995cu}
\begin{equation}
\label{eq:potential}
V(\chi) = \frac{\Mpl^2}{2 \chi^2} \left[ Q(\chi) \chi
- f(Q(\chi)) \right]
\end{equation}
Here $Q(\chi)$ solves $\chi = f'(Q)$. Finally a rescaling of the
metric (which should be thought of as a field relabeling) 
\begin{equation}
\label{eq:conf}
\tilde{g}_{\mu\nu} = \chi g_{\mu\nu} = e^{(2/\sqrt{6}) \phi/\Mpl}
g_{\mu\nu}
\end{equation}
reveals the kinetic terms for the scalar field:
\begin{eqnarray}
\label{eq:efaction}
S_{EF} &=& \int d^4x \sqrt{-\tilde{g}} \left( 
\frac{\Mpl^{2}}{2}\tilde{R}
- \frac{1}{2}\tilde{g}^{\mu\nu} \partial_\mu \phi \partial_\nu \phi
- V(\phi) \right) \nonumber \\
& & + S_M [\tilde{g}_{\mu\nu} e^{-\sqrt{\frac{2}{3}}
\frac{\phi}{M_{\rm{pl}} } }, \Psi, A_\alpha, \ldots]\,,
\end{eqnarray}
where the new canonical scalar field $\phi$ is related to $\chi, Q, R$
through
\begin{equation}
\label{eq:scalars}
f'(R) = f'(Q) = \chi = \exp\left(\sqrt{2/3}\phi/M_{\rm{pl}}\right)\,.
\end{equation}
As the kinetic terms for $\tilde{g}_{\mu\nu}$ and $\phi$ are
now both canonical, we see that these are the true degrees
of freedom of $f(R)$ gravity.
This demonstrates that the theories defined by $S_{JF}$ (
the Jordan Frame) and
$S_{EF}$ (the Einstein Frame) 
are completely equivalent when $f''(Q) \neq 0$. We choose
to analyze the theory in the Einstein Frame 
as things are much simpler here. It is,
however, important to be careful to interpret results correctly,
making reference to what is observed. In particular, matter
is defined in the Jordan Frame, and hence it will be most sensible to talk about
JF observables. We will give a simple example of this when we 
have introduced some matter.

The equations of motion for $\phi$ 
resulting from Eq.~(\ref{eq:efaction}) are 
\begin{equation}
\label{eq:scalar-eom}
-\tilde{\Box} \phi = - \frac{d V}{d \phi} 
- \frac{\tilde{T}^M}{\sqrt{6} M_{\rm{pl}} }\,,
\end{equation}
and for the metric $\tilde{g}_{\mu\nu}$,
\begin{equation}
\tilde{R}_{\mu\nu} - \frac{1}{2} \tilde{g}_{\mu\nu} \tilde{R}
= M_{\rm{pl}}^{-2} \left( \tilde{T}^M_{\mu\nu} + \tilde{T}^\phi_{\mu\nu}
\right) \end{equation}
with the energy momentum tensors
\begin{eqnarray}
\label{eq:em-einstein}
\tilde{T}^M_{\mu\nu} &=&  \chi^{-1} 
T^M_{\mu\nu} \left(\chi^{-1} \tilde{g}_{\mu\nu} \ldots \right) \\
\label{eq:em-phi}
\tilde{T}^\phi_{\mu\nu} &=& \partial_{\mu} \phi \partial_\nu \phi
+ \tilde{g}_{\mu\nu} \left( - \frac{1}{2} \tilde{g}^{\alpha\beta}
\partial_\alpha \phi \partial_\beta \phi
+ V(\phi)
\right)
\end{eqnarray} 
Note that only the combination 
$\tilde{T}^M_{\mu\nu} + \tilde{T}^\phi_{\mu\nu}$
is conserved in the EF.

There are two important observations to be made about Eq.~(\ref{eq:efaction})
relating to the extra coupling to matter. First, the 
$\tilde{T}^M/\Mpl \sqrt{6}$ term in Eq.~(\ref{eq:scalar-eom}) 
represents an additional density dependent
``force'' on the scalar field, and for special cases where
we can solve for the $\phi$ dependence of 
$\tilde{T}^M /\Mpl \sqrt{6}$ explicitly, as in
\cite{Khoury:2003rn}, we can think of the
scalar field living in an effective potential. We will see
two examples where this force is important, the most dramatic 
being the Chameleon effect.

Second, $\phi$ couples to matter as strongly as conventional
gravity ($\tilde{g}_{\mu\nu}$) does.
Hence,
as was already mentioned, $\phi$ will mediate a 
detectable fifth force 
for solar system tests
unless we do something dramatic to hide it. Finding theories which
hide $\phi$ from solar system tests is the focus of this paper.

\subsection{Matter and Cosmology in $f(R)$ theories}

Let us first consider the coupling to standard model fields,
assuming that they are defined in the JF.
This is important for understanding how $\phi$ may decay. 
Massless scalar fields conformally coupled to gravity and 
massless gauge bosons
behave the same in the two frames and so do not couple to $\phi$.
However, a minimally coupled (real) scalar field $\Phi$ and a Dirac field $\Psi$
have extra interactions with $\phi$ in the EF:
\begin{eqnarray}
S_{\Phi} &=& \int d^4x \sqrt{-\tilde{g}} \left\{
-\frac{1}{2} \left(\partial \tilde{\Phi}\right)^2
- \frac{1}{2} m_\Phi^2 \chi^{-1} \tilde{\Phi}^2  \right. \\
& & \left. - \frac{1}{12 \Mpl^2} \tilde \Phi^2 \tilde{g}^{\mu\nu}
\partial_\mu \phi
\partial_\nu \phi 
- \frac{1}{\sqrt{6} \Mpl} \tilde \Phi \tilde{g}^{\mu\nu}
 \partial_\mu \tilde{\Phi} \partial_\nu \phi \right\} \nonumber \\
\label{eq:einstein-dirac}
S_{\Psi} &=& \int d^4x \sqrt{-\tilde{g}} \bar{\tilde{\Psi}}
\left( i \tilde{\gamma}^\mu \tilde{D}_\mu
- m_\Psi \chi^{-1/2} \right)
\tilde{\Psi}\,,
\end{eqnarray}
where the JF fields have been rescaled as $\tilde \Phi = \chi^{-1/2} \Phi$
and $\tilde \Psi = \chi^{-3/4} \Psi$. Note 
that the cosmologically evolving field $\phi = \bar{\phi}(t)$
will change the masses of the standard model particles in the EF as
\begin{equation}
\label{eq:masses-scale}
\tilde{m} = m \chi^{-1/2}
\end{equation}
and small excitations $\delta \phi$ 
around the average value $\bar{\phi}(t)$ will roughly speaking
interact via the vertices defined by the interaction Lagrangian,
\begin{equation}
\label{eq:vertices}
\frac{1}{\sqrt{6} \Mpl} \left(
\tilde{m}_\Phi^2  \delta \phi \tilde{\Phi}^2
- \tilde \Phi \tilde{g}^{\mu\nu}
\partial_\mu \tilde{\Phi} \partial_\nu \delta \phi 
+ \tilde{m}_\Psi \delta \phi \bar{\tilde{\Psi}}
\tilde{\Psi} \right)
\end{equation}
to lowest order in $1/\Mpl$.
The mass shift in Eq.~(\ref{eq:masses-scale}) has an interesting
consequence in the EF; it shifts the frequency of the
absorption and emission lines by a factor of $\chi^{-1/2}$.
This effect will be indistinguishable from the normal cosmological
redshift due to expansion, and our effective redshift will be the combination of
both cosmological expansion and mass shift: $(1+z)^{-1} = \tilde{a}
\chi^{-1/2}$, where $\tilde a$ is the scale
factor in the EF normalized equal to unity today. This combination 
turns out to be the Jordan frame scale factor $a$ (see below), so
our redshift measurements coincide in both frames as expected.

Perfect fluids are best examined in the JF, because it is here
that their energy momentum tensor is conserved. For a general
JF metric one can solve for the flow of the fluid using conservation
of $T^M_{\mu\nu}$ and number flux
$n U^\mu$ (or other relevant physical principles)
and then map into the EF via Eq.~(\ref{eq:em-einstein}).

\subsubsection{The homogeneous and isotropic case}

For example, consider a homogeneous isotropic cosmology,
\begin{eqnarray}
\rm{(JF)} \quad ds^2 = d t^2 - a(t)^2 d \Vec{x}^2 \,,
&\quad & U^\mu = (\partial_t)^\mu \,, \\
\rm{(EF)} \quad d\tilde{s}^2 = d \tilde{t}^2 - 
\tilde{a}(\tilde{t})^2 d \Vec{x}^2 \,,
&\quad & \tilde{U}^\mu = \left(\partial_{\tilde{t}}\right)^\mu \,,
\end{eqnarray}
where $U^\mu$ and $\tilde{U}^\mu$ are the local fluid velocities
in the two frames. The quantities above are related by 
\begin{equation}
\tilde{a} = \chi^{\frac{1}{2}} a\,, \quad 
\mathtt{d}\tilde{t} = \chi^{\frac{1}{2}} \mathtt{d}t\,, \quad 
\tilde{U}^\mu = \chi^{-1/2} U^\mu\,.
\end{equation}
These relations imply that
\begin{equation}
\label{eq:hubble}
H = \chi^{1/2} \left( \tilde H - \frac{\tilde{\partial}_t \phi}{\sqrt{6} \Mpl}
\right)\,.
\end{equation}
For example, applying the principles of entropy ``conservation''
and number conservation in the JF
(one may also need to demand thermal and chemical
equilibrium as relevant to the early universe) results
in known $\rho(a)$ and $p(a)$ such that the EF energy
momentum tensor may be written as
\begin{equation}
\tilde{T}^M_{\mu\nu} = \tilde{\rho} \tilde{U}_\mu \tilde{U}_\nu + \tilde{p}
\left( \tilde{U}_\mu \tilde{U}_\nu + \tilde{g}_{\mu\nu} \right)\,,
\end{equation}
where 
\begin{equation}
\label{eq:densitymap}
\tilde{\rho} = \chi^{-2} \rho\left( \tilde a \chi^{-1/2}\right)\,,
\quad
\tilde{p} = \chi^{-2} p\left( \tilde a \chi^{-1/2}\right)\,.
\end{equation}
The cosmological equations of motion are,
\begin{equation}
\label{eq:einstein-cosmo}
3 \tilde{H}^2 \Mpl^2 =  \tilde{\rho}
+ \frac{1}{2} \left( \tilde{\partial}_t \phi \right)^2 + V(\phi)\,,
\end{equation}
\vspace{-12pt}
\begin{equation}
\label{eq:scalar-cosmo}
\tilde{\partial}_t^2 \phi \!+\! 3\tilde{H} \tilde{\partial}_t \phi
= - \frac{\partial V_{\rm{eff}}(\phi,\tilde a)}{\partial \phi}
= - \frac{d V_{E}}{d\phi}  - \frac{\tilde{T}^M}{\sqrt{6} \Mpl}\,.
\end{equation}
The effective potential for the scalar field coupled to 
homogeneous and isotropic matter is 
\begin{equation}
\label{eq:raweffective}
V_{\rm{eff}} (\phi,\tilde a) = V(\phi) + \tilde \rho = 
V(\phi) + \chi^{-2} \rho\left( \tilde a \chi^{-1/2}\right)\,.
\end{equation}
For the special case where the only density
components present are non-relativistic 
($\rho = \rho_{\rm{NR}} \propto a^{-3}$)
and ultra-relativistic ($\rho = \rho_{\rm{R}} \propto a^{-4}$) fluids, 
the effective potential is
\begin{equation}
\label{eq:effective}
V_{\rm{eff}}(\phi) = V(\phi) +  \bar{\rho}_{\rm{NR}}(\tilde a)
e^{-\frac{\phi}{\Mpl \sqrt{6}}}
 + \bar{\rho}_{\rm{R}}(\tilde{a})
\end{equation}
where for convenience we define 
$\bar{\rho}_{\rm{NR}}(\tilde a) \equiv \chi^{-3/2} \rho_{\rm{NR}}
(\tilde a \chi^{-1/2}) \propto \tilde a^{-3}$ and 
$\bar{\rho}_{\rm{R}}(\tilde a) \equiv \chi^{-2} 
\rho_{\rm{R}} (\tilde a \chi^{-1/2})\propto \tilde a^{-4}$.
Note that relativistic particles provide
no force on $\phi$ because $\tilde T$ vanishes, or
equivalently because $\bar{\rho}_R(\tilde a)$ appears simply as an additive
constant to the potential in Eq.~(\ref{eq:effective}).

\subsubsection{The spherically symmetric case}

We now turn to the case of a spherically symmetric distribution of 
non-relativistic matter $\rho_{\rm{NR}}(r)$ in the JF,
for which we aim to solve for the metric $g_{\mu\nu}$. We
wish to consider this problem in the EF, where $\phi$
will take a spherically symmetric form and gravity
behaves like GR coupled to $\tilde{\rho} = \chi^{-2} \rho_{\rm{NR}}$.
In the
weak field limit, we write the metrics in the two frames as
\begin{subequations}
\begin{eqnarray}
\nonumber 
\rm{(JF)}\quad ds^2 &=& - (1- 2 A(r)) d t^2 
+ (1 + 2 B(r) ) d r^2  \nonumber \\ 
\label{eq:EFmetric}
&& \hphantom{ - (1- 2 A(r)) dt^2  (1 + 2 ) }
+ r^2 d\Omega^2 \\
\rm{(EF)}\quad d\tilde{s}^2 &=& - (1- 2 \tilde{A}(\tilde r)) 
d t^2 
+ (1 + 2 \tilde{B}(\tilde r) ) d \tilde{r}^2 \nonumber \\
&& \hphantom{ - (1- 2 A(r)) d t^2  (1 + 2 ) }
+  \tilde{r}^2 d \Omega^2 
\end{eqnarray}
\end{subequations}
where $\tilde r = \chi^{1/2} r$ and for small $\phi/\Mpl$,
the gravitational potentials are related by
\begin{subequations}
\label{eq:potentials}
\begin{eqnarray}
A(r) &\approx& \tilde{A}(\tilde r) + \frac{\phi(\tilde r)}{\sqrt{6} \Mpl}\,,
\\
B(r) &\approx& \tilde{B}(\tilde r) + 
\frac{1}{\sqrt{6} \Mpl} \frac{ d \phi(\tilde r)}{d \ln \tilde r}\,.
\end{eqnarray}
\end{subequations}
Following \cite{Khoury:2003rn} we define a 
non-relativistic energy density
$\bar{\rho}_{\rm{NR}}(\tilde r)  
= \chi^{-3/2} \rho (r)$ in the EF
which is conserved there and is analogous
to $\bar{\rho}_{\rm{NR}}(\tilde a)$ defined above for cosmology. 
Ignoring the back reaction of the metric on $\phi$, we take
$\tilde{g}_{\mu\nu} \approx \eta_{\mu\nu}$
in Eq.~(\ref{eq:scalar-eom}) and find as in \cite{Khoury:2003rn} that
\begin{equation}
\label{eq:spherical}
\frac{1}{\tilde r^2} \frac{d}{d \tilde r} \left( \tilde r^2
\frac{d \phi}{ d\tilde r}\right) = V'(\phi) - 
\chi^{-\frac{1}{2}} \frac{ \bar{\rho}_{\rm{NR}}(\tilde r)}{ \sqrt{6} \Mpl}
= \frac{ \partial V_{\rm{eff}}(\phi , \tilde r)}{ \partial \phi}\,,
\end{equation}
where again the effective potential is 
\begin{equation}
\label{eq:effective2}
V_{\rm{eff}} 
= V (\phi) + \chi^{-1/2} \bar{\rho}_{\rm{NR}}(\tilde r)\,.
\end{equation}
Solving Eq.~(\ref{eq:spherical}) for $\phi$ then allows us to find the
metric in the JF via Eq.~(\ref{eq:potentials}).

As an instructive example, consider the quadratic 
potential $V(\phi) = m_\phi^2 \phi^2/2$
and a uniform sphere of mass $M_c$ and radius $R_c$,
the solution external to 
the sphere is given by a Yukawa potential
\begin{equation}
\label{eq:yukawa}
\frac{\phi(r)}{\Mpl} = \frac{1}{\sqrt{6}} \frac{ M_c e^{-m_\phi r}}
{ 4 \pi \Mpl^2 r}
\end{equation}
assuming that $m_\phi R \ll 1$ and $\phi/\Mpl \ll 1$ so that
$\tilde r \approx r$.
If we ignore the energy density of the profile $\phi(r)$,
then outside the object there is vacuum.
The metric in the EF is then simply the Schwarzschild solution for
mass $M_c$. In other words, the potentials 
in Eq.~(\ref{eq:EFmetric}) are given by
$\tilde A(\tilde r)  = \tilde B(\tilde r) = M_c / 8 \pi \Mpl^2 \tilde r $
in the weak field limit $|\tilde A|,\, |\tilde B| \ll 1$.
In the JF using Eq.~(\ref{eq:potentials}), one finds the corresponding
potentials
\begin{subequations}
\begin{eqnarray}
A(r) &\approx& \tilde{A}(r) \left[ 1 + \frac{1}{3} e^{-m_\phi r} \right]\,, \\
B(r) &\approx& \tilde{A}(r) \left[1 - \frac{1}{3} e^{-m_\phi r}
(1 + m_\phi r ) \right]\,.
\end{eqnarray}
\end{subequations}
For $r \ll m_\phi^{-1}$ we find that the PPN parameter $\gamma=1/2$,
a well known result for a Brans Dicke theory \cite{Brans:1961sx}
with $\omega_{BD} = 0$ \cite{Will:1993ns}. 

The key feature here is the effective potential
from Eqs.~(\ref{eq:effective}) and (\ref{eq:effective2}). 
We have now seen that it makes a crucial difference
in two situations, and it will
play an important role in the next two sections as well.

\section{An $f(R)$ Chameleon \label{sec:cham}}

In this section, we consider $f(R)$ theories that are able
to pass solar system tests of gravity because of the 
so-called ``Chameleon'' effect.
We first present a theory that
is by design very similar to the original Chameleon 
model presented in \cite{Khoury:2003aq}.
We will give a brief description of how this model
evades solar system constraints,
and then move on to the cosmology of these $f(R)$ theories,
concentrating in particular on their relation to DE. 
Throughout this discussion we refer the
reader to the original work \cite{Khoury:2003aq,Brax:2004qh,Khoury:2003rn},
highlighting the differences between the original
and $f(R)$ Chameleons.

The Chameleon model belongs to the following general class of models,
\begin{equation}
f(R) = R - (1-m) \mu^2 \left(\frac{R}{\mu^2}\right)^m - 2 \Lambda.
\end{equation}
The sign of the second factor is important to
reproduce the Chameleon,
and the $(1-m)$ factor ensures that the theory is equivalent
to GR as $m\rightarrow 1$.
These models have been considered before in the literature
\cite{Carroll:2003wy,Amendola:2006kh}; 
in particular, this class contains the original DE $f(R)$ of
Eq.~(\ref{eq:arch}) 
when $m=-1$ and $\Lambda=0$.

The potential for $\phi$ in the EF is
\begin{equation}
\label{eq:expansion}
V(\phi) = \frac{\Mpl^2 \mu^2}{2 \chi^2} (m-1)^2 \left(
\frac{\chi-1}{m^2-m} \right)^{\frac{m}{m-1}} + \frac{\Mpl^2 \Lambda}{\chi^2}.
\end{equation}
where $\chi = \exp( \sqrt{2/3} \phi/\Mpl)$ as usual.
For $0<m<1$ and for $|\phi/\Mpl| \ll 1$, this reduces to
\begin{equation}
\label{eq:champot}
V_{E}(\phi) = M^{4+n} \left(-\phi\right)^{-n} + \Mpl^2 \Lambda,
\end{equation}
defined for $\phi < 0$, 
where the old parameters $\mu,m$ are related to the new
parameters $M,n$ through
\begin{equation}
m = \frac{n}{1+n} ,\quad
\mu^2 = 
\frac{ \left( 2 (1+n)^2 \right)^{1+n} }{\left( \sqrt{6} n \right)^n}
\frac{ M^{4+n} }{\Mpl^{n+2}}\,.
\end{equation}
The preferred values used in \cite{Khoury:2003aq}
are $M \sim 10^{-3} \rm{eV}$ and $n \sim  1$.
In the $f(R)$ theory, these values
give $m \sim 1/2$ and $\mu \sim 10^{-50} \rm{eV}$, i.e. 
much smaller than the Hubble scale today.

For small $|\phi|/\Mpl$, this
singular potential is equivalent
to the potential considered in \cite{Khoury:2003aq} 
for the Chameleon scalar field, albeit
with $\phi \rightarrow -\phi$. The coupling to matter, which
is a very important feature of this model, is also
very similar.  In \cite{Khoury:2003aq}, a species of particles $i$ is
assumed to have its own Jordan Frame metric $g^{(i)}_{\mu\nu}$,
with respect to which it is defined
and a conformal coupling to the metric in the EF 
\begin{equation}
\label{eq:beta}
g^{(i)}_{\mu\nu} = e^{2\beta_i \phi/\Mpl}\tilde{g}_{\mu\nu}.
\end{equation}
Comparing this to Eq.~(\ref{eq:conf}),
the $f(R)$ Chameleon has $\beta_i = - 1/\sqrt{6}$ for all
matter species, so that all the Jordan Frame metrics coincide.

In the original Chameleon model, the $\beta_i$ were specifically
chosen to be different so that $\phi$ would show up in tests
of the weak equivalence principle (WEP). The $f(R)$ Chameleon does
not show up in tests of the WEP, so the
solar system constraints will be less stringent here.

\begin{figure}
\psfig{figure=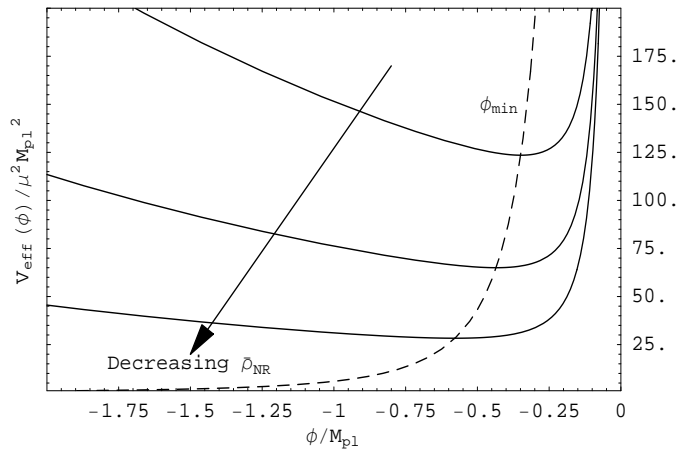,width=9cm,height=6cm}
\caption{Effective potential for the Chameleon model Eq.~(\ref{eq:expansion})
with decreasing $\bar{\rho}_{\rm{NR}}/\mu^2 \Mpl^2 = 100,50,20
$ and $m=1$. Note that $\phi_{\rm{min}}$ and the mass $m_\phi^2$
(the curvature of the minimum)
are very sensitive to the background energy density
$\bar{\rho_{\rm NR}}$.\label{fig:champot}}
\end{figure}

This coupling to matter, along with the singular potential,
are the defining features of this $f(R)$ that make it a Chameleon 
theory. 
The effective potential $V_{\rm{eff}}$, discussed
in the previous section (see for example Eq.~ \ref{eq:effective}),
is then a balance between two forces;
$V$ pushing $\phi$ toward more negative values and the 
density-dependent term pushing $\phi$ toward more positive values.
So although the singular potential Eq.~(\ref{eq:champot})
has no minimum and hence no stable ``vacuum'', the effective potential
Eq.~(\ref{eq:effective}) including the coupling to matter does have a minimum. 
In fact,  
the density dependent term pushes the scalar field $\phi$ up
against the potential wall created by the singularity in
$V$ at $\phi=0$. 
Indeed, the field value $\phi_{\rm min}$ at the minimum
of the effective potential $V_{\rm{eff}}$ and the 
mass $m_\phi$ of $\phi$'s excitation around that 
given minimum are both highly
sensitive increasing functions of 
the background density $\bar{\rho}_{\rm NR}$,
as illustrated in Fig.\ \ref{fig:champot}. Using
Eq.~(\ref{eq:effective2}) for small $|\phi|/\Mpl$, the 
field value at the minimum 
and the curvature of the minimum are, respectively,
\begin{eqnarray}
\label{eq:minsolar}
- \frac{\phi_{\rm min}}{\sqrt{6} \Mpl}
&=& \frac{ m(1-m)}{2} \left( \frac{ \Mpl^2 \mu^2}
{ \bar{\rho}_{\rm{NR}}}
\right)^{1-m} \, ,  \\
\label{eq:masssolar}
m_\phi^2 &=& \frac{2}{3(1-m)} \frac{\bar{\rho}_{\rm NR}}{\Mpl^2}
\left( -\frac{\sqrt{6} \Mpl}{\phi_{\rm min}} \right)\,.
\end{eqnarray}
It is plausible that a scalar field $\phi$ which is very light for
cosmological densities is heavy for solar system densities
and hence currently undetectable. 
However, as we will now see, the actual mechanism that
``hides'' $\phi$ from solar system tests is a bit more complicated
than this.

\subsection{Solar System Tests}

In this section, we will derive solar system and laboratory
constraints on the parameters $(\mu,m)$, summarized
in Figure \ref{fig:cham}.
The profile of $\phi(\tilde r)$ in the solar system (around the Earth,
around the Sun, etc.) is of interest for solar system tests
of gravity: it determines the size of the 
fifth force and the post-Newtonian
parameter $\gamma$.
Because the effective potential for $\phi$ changes in different
density environments, the differential equation governing the
profile $\phi(\tilde r)$ in
Eq.~(\ref{eq:spherical}) is highly non-linear, and
the standard Yukawa profile of Eq.~(\ref{eq:yukawa}) does not always arise. 
These non-linear
features have been studied in \cite{Khoury:2003rn}, where
it was found that
for a spherically symmetric object of 
mass $M_c$ and radius $R_c$ surrounded by a gas of asymptotic density 
$\rho_\infty$, the profile is governed by the so-called 
``thin-shell'' parameter,
\begin{equation}
\label{eq:thinshell}
\Delta = \frac{| \phi_{\rm{min}}^\infty - \phi_{\rm{min}}^c |}{\sqrt{6} \Mpl} 
\frac{24 \pi \Mpl^2 R_c}{M_c},
\end{equation}
where $\phi_{\rm min}^\infty$ and 
$\phi_{\rm min}^c$ are the minima of the effective potential
in the presence of the asymptotic energy density 
$\bar{\rho}_{\rm{NR}} = \rho_\infty$ 
and central energy densities $\bar{\rho}_{\rm{NR}} = \rho_c$
respectively, see Eq.~(\ref{eq:minsolar}).
If $\Delta$ is large, then the external profile of $\phi$ is the
usual Yukawa profile Eq.~(\ref{eq:yukawa})
with mass $m_\infty=m_\phi~(\bar{\rho}_{\rm{NR}}=\rho_\infty)$,
the curvature of the effective potential in the 
presence of the asymptotic density; see Eq.~(\ref{eq:masssolar}).
If $\Delta$ is small, then the Yukawa profile is suppressed
by a factor of $\Delta$. The term
``thin shell'' comes from the fact that only a portion
of such a ``thin shell'' object contributes to the external Yukawa profile, 
the thickness of the shell being roughly $ (\Delta R_c)$.
We simply treat $\Delta$ as a parameter that suppresses this profile
if $\Delta \ll 1$.

For example, let us consider the profile
$\phi$ around the Sun, with $M_c = M_{\rm Sun}$ and 
$R_c = R_{\rm Sun}$. Assuming that we are in the 
thin shell regime ($\Delta\ll 1$), the Yukawa
profile of Eq.~(\ref{eq:yukawa})
suppressed by a factor $\Delta$ becomes,
\begin{equation}
\phi(r) = \frac{\Delta}{\sqrt{6}}  
\frac{M_{\rm Sun} e^{- m_\infty r} }{4 \pi \Mpl r} + \phi_{\rm min}^\infty \,.
\end{equation}
As in \cite{Khoury:2003aq}, we take the asymptotic 
density used to find $\phi_\infty$ and $m_\infty$
as that of the local homogeneous density of dark and baryonic matter
in our Galaxy: $\rho_\infty \approx  10^{-24} \rm{g}/\rm{cm}^3$.
Following the discussion in Section \ref{sec:equiv},
the metric in  the EF external to the Sun is just the Schwarzschild metric
(in the weak field limit) with
Newtonian potential $\tilde{A}(r) \approx M_{\rm Sun}/(8\pi\Mpl^2 r)$.
Using Eq.~(\ref{eq:potentials}) to map this metric 
into the JF metric 
$g_{\mu\nu} 
= \chi^{-1} \tilde{g}_{\mu\nu}$, we find
\begin{eqnarray}
\label{eq:metric-JF}
d s^2 &=& - \left[1 - 2 \tilde{A}(r) \left( 1 + \frac{\Delta}{3} e^{-m_\infty r}
) \right) \right] d t^2 
+ r^2 d\Omega^2  \nonumber
\\
 & +& \left[1 + 2 \tilde{A}(r) \left(1- \frac{\Delta}{3} e^{-m_\infty r} (
1+ m_\infty r) \right) \right] d r^2\,.
\end{eqnarray}
Assuming that the Compton wavelength $m_\infty^{-1}$ is much
larger than solar system scales (we will confirm this later),
we obtain within the PPN formalism \cite{Will:1993ns} that
\begin{equation}
\gamma = \frac{3-\Delta}{3+\Delta}
\approx 1 - (2/3) \Delta
\end{equation}
There are several observational constraints
on $|\gamma-1|$, including ones from the deflection of light
and from Shapiro time delay. The tightest solar system
constraint comes from Cassini 
tracking, giving $|\gamma-1| \lesssim 2.3 \times 10^{-5}$ \cite{Will:2005va}.
Thus the  ``thin shell'' parameter satisfies
$\Delta \lesssim 3.5 \times 10^{-5}$. We take
$|\phi_{\rm min}^c| \ll |\phi_{\rm min}^\infty|$ because of the
sensitive dependence on the local density, so
the definition of $\Delta$ in Eq.~(\ref{eq:thinshell}) becomes
\begin{equation}
\Delta \approx 3 |\phi_{\rm min}^\infty|/ \sqrt{6} \Mpl \tilde{A}(R_{\rm Sun})
\end{equation}
where $\tilde{A}(R_c = R_{\rm{Sun}})
\approx 10^{-6}$ is the Newtonian potential at the surface
of the Sun. Using Eq.~(\ref{eq:minsolar})
with 
$\bar{\rho}_{\rm{NR}} = \rho_\infty \approx  10^{-24} \rm{g}/\rm{cm}^3$ 
gives the constraint
\begin{equation}
\label{eq:sol}
\frac{\mu^2}{H_0^2} \lesssim  3 \left( \frac{2}{m(1-m)} \right)^{\frac{1}{1-m}}
10^{\frac{-6 -5 m}{1-m} }
\end{equation}
on the theory parameters $\mu$ and $m$.
For theories which fail this bound, we find that the 
Compton wavelength of
$\phi$ for the asymptotic background density of our galaxy 
satisfies
$m^{-1}_\infty \gtrsim 10^{10} \rm{AU}$. This confirms
the assumption that $m^{-1}_\infty$ is large compared
to solar system scales, which was used to derive this bound.

As was already noted, the solar system constraints
derived in \cite{Khoury:2003rn} are more restrictive. This is
because they demanded that the couplings ($\beta_i$) to different
species of particles in equation (\ref{eq:beta}) be different.
This
gives violations of the weak equivalence principle
on Earth-based experiments unless the Earth and
atmosphere have a thin shell.  
However, in the $f(R)$ Chameleon model, all the $\beta_i$
are the same, so there will be no 
weak equivalence principle violations. 

The $f(R)$ Chameleon may still show up in searches
for a fifth force, in particular in tests of the inverse
square law. 
The strongest comes from Earth-based
laboratory tests of gravity such as in the E\"{o}t-Wash 
experiments \cite{Hoyle:2000cv}. 
By demanding that the test masses acquire thin shells, 
\cite{Khoury:2003rn} 
found constraints on the parameters $(M,n)$ which map
into the following bound on the
$f(R)$-parameters $(\mu,m)$:
\begin{equation}
\label{eq:ssconstraints}
\frac{\mu^2}{H_0^2} \lesssim
(1-m) \left(\frac{2}{m(1-m)}\right)^{\frac{m}{1-m}} 
10^{\frac{-4 -24 m}{1-m}}
\end{equation}

\subsection{Cosmology}

We now turn to
the cosmology of the Chameleon scalar field, which
was studied in \cite{Brax:2004qh}.  It was found there and already commented
on in \cite{Khoury:2003rn} that the mass of $\phi$ on cosmological
scales is not small enough to give any interesting DE behavior
for $M \approx 10^{-3} \rm{eV}$ and $n \sim 1$. We will
revisit this question in the $f(R)$ context: do any
\emph{allowed} parameters ($\mu,m$) in 
Eq.~(\ref{eq:ssconstraints}) give non-vanilla DE?
Will there be any cosmologically observable differences between
this $f(R)$ Chameleon and the original model (which
is in principle possible because higher
order terms in the expansion of $V$ in Eq.~(\ref{eq:expansion})
may become important)? 
We will see that the answer to both of these questions
is \emph{no} for the same reason: solar system tests
preclude the minimum of the effective potential from 
lying at $\phi \lesssim - \Mpl$ on cosmological scales today.

Let us try to understand this by looking at the details
of Chameleon cosmology. We first note that, as opposed to
\cite{Brax:2004qh}, we do \emph{not} fix $\Lambda \Mpl^2 = M^4$,
so we are less restricted as to what $M$ can be. 
The essence
of the argument, however, is the same as in \cite{Brax:2004qh}.
Working in the
EF, for a large
set of initial conditions in the early universe, $\phi$ is attracted 
to the minimum of the effective
potential given by Eq.~(\ref{eq:effective}). 
The scalar field tracks the minimum, which shifts 
$\phi(\tilde a) \equiv \phi_{\rm min}$ as the universe expands. The 
energy density in coherent oscillations around this minimum
are negligible and so there is no ``moduli problem''. (
In contrast, this \emph{may} be a problem
for the case considered below in Section \ref{sec:mass}.)  

We will see that the condition for such a tracking solution
to be valid is that the minimum satisfies
\begin{equation}
\label{eq:less}
-\phi(\tilde a)/\Mpl \ll 1,
\end{equation}
so we consistently make this assumption to derive properties
of the tracking minimum. 
After matter-radiation equality we have the tracking solution
\begin{equation}
- \frac{\phi(\tilde a)}{\sqrt{6} \Mpl}
= \frac{ m(1-m)}{2} \left( \frac{ \Mpl^2 \mu^2}
{ \bar{\rho}_{\rm{NR}}(\tilde a) + 4 V(\phi(\tilde a)) }
\right)^{1-m} .
\end{equation}
Along this tracking solution, the curvature (mass)
around the minimum and the speed of the minimum
are, respectively
\begin{equation} 
\label{eq:mass}
\frac{m_\phi^2(\tilde a)}{\tilde{H}^2}
= \frac{2}{1-m} \left(\frac{\sqrt{6} \Mpl}{-\phi(\tilde a)}\right)
\left( \frac{ \bar{\rho}_{\rm{NR}}(\tilde a) + 4 V(\phi(\tilde a)) }
{\bar{\rho}_{\rm{NR}}(\tilde a) + V(\phi(\tilde a)) } \right)\,,
\end{equation}
\begin{equation}
\frac{- 1}{\Mpl \tilde H} \frac{ d\phi(\tilde a)}{d \tilde t}
= -3 \left( \frac{- \phi(\tilde a)}{\Mpl} \right)
\frac{(1-m) \bar{\rho}_{\rm{NR}}(\tilde a)}{ \bar{\rho}_{\rm{NR}}(\tilde a) + 4 V(\phi(\tilde a)) }
\end{equation}
Since $\phi$ will track the minimum while $m_\phi(\tilde a) \gg \tilde H$,
Eq.~(\ref{eq:mass}) shows that the assumption of Eq.~(\ref{eq:less}) 
is indeed consistent. 

\begin{figure}[h!]
\hbox{
\hspace{-.5cm}
\psfig{figure=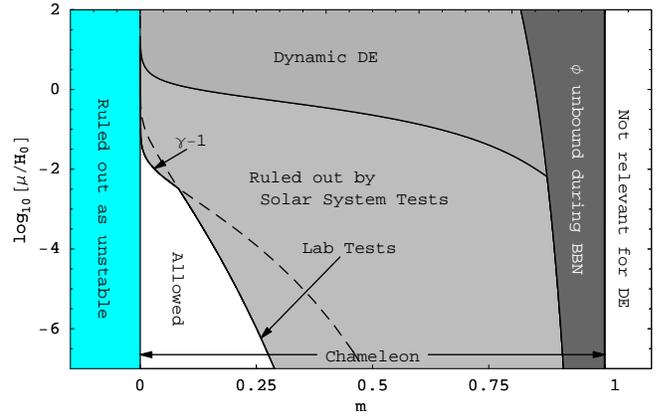,width=9cm}}
\caption{Solar system constraints on the $f(R)$ Chameleon are seen to exclude 
all models where the ``dark energy'' is observationally distinguishable from a cosmological constant
(labeled ``dynamic DE''). 
The two different solar system constraint curves come
from Eq.~(\ref{eq:sol}) and Eq.~(\ref{eq:ssconstraints}). Although it is not
clear from the plot, the limits $m\rightarrow 0$, $m\rightarrow1$
and $\mu \rightarrow 0$ are all acceptable and yet give
no dynamical DE. Indeed these are exactly the limits in which
we recover standard GR.
\label{fig:cham}}
\end{figure}

During radiation domination, $m_\phi^2(\tilde a)/\tilde{H}^2 \sim 
(-\Mpl/\phi(\tilde a)) \, \tilde a / \tilde a_{\rm{MR}}$, so 
it is possible that at early times the scalar field is unbound
at early times.
We know the expansion history
and the effective value of Newtons constant $G_N$ quite well \cite{Carroll:2001bv,Copi:2003xd}
around BBN; 
if $\phi$ is 
unbound, we have no reason to believe that $G_N$ is near
today's value.
Requiring that it is bound before the beginning of BBN gives a 
constraint that we have included in Figure \ref{fig:cham}.

Returning to the matter-dominated era, Eq.~(\ref{eq:less})
implies that the 
expansion history in the JF may be written as
\begin{equation}
\label{eq:freed}
3 \Mpl^2 H^2 = \rho_{\rm{NR}}(a) + V(\phi(a_0)) + 
\mathcal{O}(\frac{\phi}{\Mpl})\,,
\end{equation}
For $|\phi (a_0)|/\Mpl \ll 1$ 
today, this is just the usual Friedmann equation
with a cosmological constant,
where in accordance with experiment 
we are forced to identify $V(\phi(a_0))$ with
$\rho_X(0)$, the current dark energy
density. 
Note that the parameter
$\Lambda$ in $V$, which we have not fixed,
allows us to make this choice independent
of any values of $\mu$ and $m$. For $m$ not small, $
\Lambda \Mpl^2 \approx \rho_X(0)$;
however, for small $m$ we will see later that the situation
will be slightly different.

This implies that the only
way to get interesting late-time cosmological behavior
is to not have $|\phi (a_0)/\Mpl| \ll 1$ but
rather $|\phi (a_0)/\Mpl| \sim 1$
today. In this case the tracking solution above is not valid; the scalar
field is no longer stuck at the minimum, and we might not have
to invoke a constant  
$\Lambda$ in $V$ to explain todays accelerated expansion.  
Rather the acceleration would be driven by 
a quintessence type phase.

However, one can show that given the solar system
constraints,
$|\phi(a_0)/\Mpl| \sim 1$ is not possible.
In fact, as we will now show, a stronger statement can be
made: even if we continue
to assume Eq.~(\ref{eq:less}), so that the tracker solution is still
valid, the solutions that are consistent with solar system
tests always give DE behavior that is ``vanilla,'' i.e.,
indistinguishable from a cosmological constant.

In these models, the
effective dark energy density is
\begin{eqnarray}
\nonumber
\rho_X(a) &\approx& V(\phi(a)) + \left(\frac{- \phi(a)}{\sqrt{6}\Mpl}\right)
\left( \rho_{\rm{NR}}(a) + \rho_X(0) \right)  \\
\label{eq:effcc}
&&\hphantom{\left(\frac{- \phi(a)}{\sqrt{6}\Mpl}\right)}
\times \left( 2 + \frac{ 6 \rho_{\rm{NR}} (a) (1-m)}
{\rho_{\rm{NR}}(a) + 4 \rho_X(0)} \right),
\end{eqnarray}
where $V(\phi(a)) - \rho_X(0)= \mathcal{O}(\phi/\Mpl)$. 
If we expected Eq.~(\ref{eq:effcc}) to give interesting behavior
in the allowed region of parameter space, we would fit 
the Friedmann equation with $\rho_X(a)$ to the combined knowledge
of the expansion history and find the allowed values of
$(\mu,m)$. We will instead adopt a simpler
approach, \emph{defining} ``non-vanilla
DE'' through the effective equation of state parameter,
\begin{equation}
w_X = -\frac{1}{3} \frac{d \ln \rho_X(a) }{d \ln a} -1.
\end{equation}
This is the relevant equation of state that one would
measure from the expansion history (that is \emph{not} 
$p_\phi / \rho_\phi$).
We say that the DE is non-vanilla if
$|w_X+1|>.01$, which is quite-optimistic
as to future observational capabilities 
\cite{Bock:2006yf}.  However, because
our result is
null the exact criterion is not important. 

The resulting constraint
on $\mu$ and $m$ is shown in Figure \ref{fig:cham} along with
the solar system constraints. As the Figure shows, all models
consistent with solar system tests are ``vanilla'' -- that is, 
indistinguishable from a cosmological constant.

The most interesting
part of this parameter space is the limit $m\rightarrow 0$,
which is one of the limits in which we
should recover general relativity.  The theory then becomes
\begin{equation}
f(R) \approx R - (\mu^2 + 2 \Lambda ) + \mu^2 m \ln \left( R /\mu^2 \right) 
\end{equation}
with the Chameleon-like (singular at $\phi=0$) potential 
\begin{equation}
V (\phi) \approx 
\frac{\Mpl^2}{2} e^{-\frac{4}{\sqrt{6}} \frac{\phi}{\Mpl} }
\left( \mu^2 + 2 \Lambda -  m \mu^2 \ln\left(1 - e^{\sqrt{\frac{2}{3}} 
\frac{\phi}{\Mpl}} \right) \right)
\end{equation}
In this limit we 
are forced to fix $(\mu^2/2 + \Lambda) \Mpl^2 \equiv \rho_X(0)$.
for different values of $\mu$. 
The DE energy equation of state parameter is
$\omega_X = -1 -0.05 m \mu^2 / H_0^2$.
The tightest solar system constraint on $\mu^2$ in this limit
is from $|\gamma-1|$ in Eq.~(\ref{eq:sol}) which gives 
$m \mu^2 \lesssim 6 \times 10^{-6} H_0^2$.   The equation
of state parameter for DE is then constrained to be $|\omega_X+1|
\lesssim 0.3 \times 10^{-6}$ which is definitely unobservable. 

Finally we note that the ultimate fate of the $f(R)$ chameleon
is different from that of the original model. This is because
$V(\phi)$ actually does have a minimum relevant for
cosmological energy densities. This is due to the
$\phi$ dependence of the
$\Lambda \Mpl^2 \chi^{-2}$ term in 
Eq.~(\ref{eq:expansion}), which is absent in the
original models. Eventually $\phi$ will settle into this minimum
and the universe will enter an inflating de Sitter phase, much like
the fate of a universe with a simple cosmological constant.
The original model on the other hand \emph{eventually} enters
a quintessence like expansion. However, this distinction
is unobservable today.

In conclusion to this section, we 
have found a previously unstudied class of $f(R)$ theories
that gives acceptable local gravity by exploiting the
Chameleon effect. For the allowed
parameters of this model, there is no interesting late-time
cosmological behavior (observably dynamic DE). 
That is not to say that these
models have no interesting physics --- there may 
indeed be some interesting
effects of such models for future solar system tests
\cite{Khoury:2003rn} or on large scale structure \cite{Brax:2005ew},
and this might warrant further study in the context of $f(R)$ models.
We also noted that the $f(R)$ model is
subtly different from the original Chameleon model.
It does not violate the weak equivalence principle
, so solar system constraints are less
stringent and the ultimate fate of the universe is now simply an 
inflating de Sitter spacetime.

This mechanism might also be a starting point for constructing
working modified gravity models which do give non-vanilla DE,
somehow exploiting this mechanism more effectively and to bridge
the gap in Figure \ref{fig:cham}
between solar system constraints and non-vanilla DE.
We suspect they will not be as simple as the one presented. This
mechanism may also be relevant for attempting to understand
the Newtonian limit of the artificially constructed $f(R)$ models
that reproduce an exact expansion history. We make this claim because
an important property of the model presented in \cite{Song:2006ej}
is that the parameter $B \propto f''(R)$ is a rapidly growing function of
the scale factor $a$. 
For small $f''(R)$, one can show that 
the mass curvature of $V$ 
is $m_\phi^2 \sim 1/f''(R)$.
Hence, in this theory the mass of the 
scalar field during cosmological evolution 
is large at early times and small at late times, as in
the Chameleon models. A more detailed
analysis, beyond the scope of this paper, is required
to see whether non-linear effects play a part in the Newtonian limit
of these theories.

\section{Massive $f(R)$ theories \label{sec:mass}}

We now consider arguably
more realistic $f(R)$ theories, namely
polynomials 
$f(R) =-2 \Lambda + R + a R^2 + b R^3  \ldots$. 
These theories have been extensively studied, especially
for quadratic $f(R)$; see \cite{Schmidt:5} and references therein.
They are more natural from the point of view
of renormalization and effective field theories: a high energy
completion of gravity would allow us to find these higher order terms.
However, common wisdom would have the higher order terms suppressed by 
inverse powers of $\Mpl$ and would force us to include other
terms of the same mass dimension such as $R^{\mu\nu} R_{\mu\nu}$. 
Despite this, we wish to explore the phenomenology
of such polynomial $f(R)$ theories and hence constrain them
with cosmological observations. In doing so, we will 
explore the full range of values for the coefficients ($a,b,...$) 
of the higher order terms to
be conservative rather than assume that they are order unity in Planck units. 

This class of theories can only match the 
the currently observed cosmic acceleration via an explicit cosmological constant term 
$f(0)$, giving the identification
$\Lambda = 3 H_0^2 \Omega_\Lambda$,
so there is 
no hope of dynamical DE.
Rather, these theories are more relevant to
very early universe cosmology where $R$ is large, and hence some of our results will be
quite speculative. 

\begin{figure}[h!]
\hbox{\psfig{figure=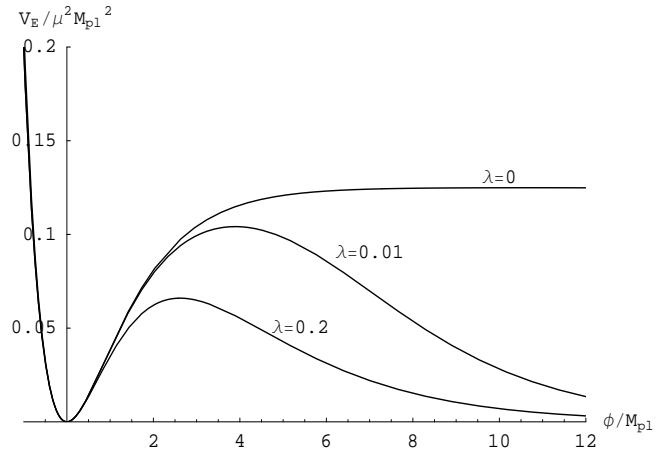,width=9cm,height=6cm}}
\caption{Potential for the $f(R)$ model in
Eq.~(\ref{eq:polymodel}) with various values of $\lambda$. Notice how
the $\lambda=0$ case has an asymptotically flat potential as
$\phi\rightarrow\infty$.
\label{fig:polymodel}}
\end{figure}

Consider for simplicity the two-parameter model
\begin{equation}
\label{eq:polymodel}
f(R) = R + R \left(\frac{R}{\mu^2} \right) +  \lambda R
\left(\frac{R}{\mu^2} \right)^2
\end{equation}
We restrict to the parameter range $\mu^2 >0$ and $0 < \lambda < 1/3$,
so that the resulting potential $V$ has a stable quadratic minimum 
and is defined for all $\phi$.
The Einstein frame potential for $\phi$ or $\chi$ is given by
\begin{equation}
\label{eq:polypot}
V_{E}(\chi) = \frac{ \Mpl^2 \mu^2}{2 \chi^2} q^2 \left( 1 + 2 \lambda q\right),
\end{equation}
where 
\begin{equation}
\label{eq:qDefEq}
q \equiv {1\over 3\lambda}\left[\sqrt{1 - 3\lambda(1-\chi)}- 1\right]
\end{equation}
is the larger of the two roots of $1-\chi + 2q + 3 \lambda q^2$  
(this ensures that the
resulting potential has a stable minimum). 
We plot this
potential for various $\lambda$ in Figure \ref{fig:polymodel}.

We will first explore the possibility that $\phi$ is the inflaton,
then discuss other constraints from our knowledge about
the early universe. Figure~\ref{fig:polycons}
summarizes our constraints.

\subsection{$f(R)$ inflation}

\begin{figure}[h!]
\hbox{\hspace{-.3cm}\psfig{figure=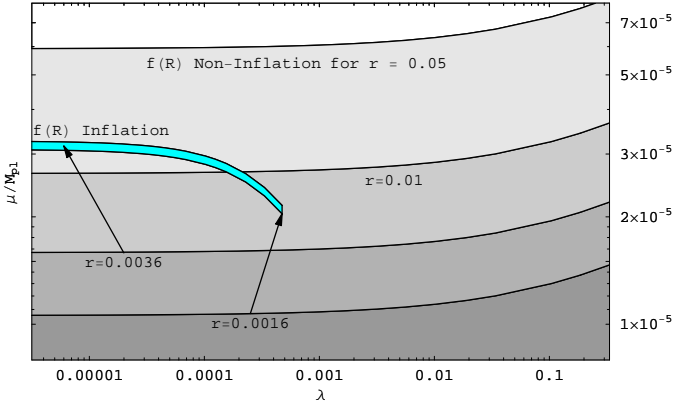,width=9cm}}
\caption{Constraints on the cubic $f(R)$ model. The thin
blue/grey sliver correspond to observationally allowed
$f(R)$ inflationary scenarios. 
Shaded are regions
we may rule out given a measurement of the tensor to scalar
ratio $r$ and the assumption that they were generated
by a period of slow roll
inflation in the early universe. The $r=0.05$ and $r=0.01$ are the
most realistic curve, in the sense that future experiments
are sensitive to such values as low as $r=0.01$ \cite{Bock:2006yf}.
\label{fig:polycons}}
\end{figure}

The possibility that higher order corrections
to the gravitational Lagrangian might be responsible for a de Sitter inflationary
period was pointed out early on in the inflationary game \cite{Starobinsky:1980te}.
For $\lambda = 0$, the potential $V(\phi)$ is  very flat
for large $\phi$, which is perfect for inflation.
This model was extensively studied in \cite{Mijic:1986iv,Mukhanov:1989rq,Hwang:2001pu},
which did find a viable inflationary model. 
We now search for
possible inflationary scenarios with $\lambda \neq 0$ that are consistent with
current observations. This
question was already considered in \cite{Berkin:1990nu}, which
found $\lambda \ll 1$; however, we wish to be more quantitative
in light of the latest CMB measurements. 

As usual in these models, it is important to keep careful
track of whether we are working in the EF or the JF: recall
that the potential
$V$ is defined in the EF, while matter is most naturally 
considered in the JF. Nonetheless,
we will argue that the inflationary
predictions are exactly the same as those of general relativity 
plus a normal
slow rolling scalar field with potential $V(\phi)$. The argument
goes as follows. Slow roll inflation works normally in the EF where
the graviton and scalar field have canonical actions. In
particular, the EF is where one \emph{should} calculate the spectrum
of tensor and scalar mode fluctuations. Re-heating
and the transformation of fluctuations in $\phi$ to 
adiabatic density fluctuations also works as usual in the EF, because at this time
the cosmic fluid is relativistic and hence governed by the same 
equations of motion in both frames. After reheating, $\phi$
is frozen out at the minimum of $V$, 
with $\phi = 0$ and $\chi=1$, so 
there is no longer any distinction
between the JF and the EF ($\tilde{g}_{\mu\nu} = g_{\mu\nu}$). 
Calculations for $\lambda=0$ were performed both as above
and in the JF in \cite{Hwang:2001pu}, and the results were
found to be consistent as expected.

Using this idea, calculating the inflationary predictions is
straightforward.  Using Eq.~(\ref{eq:vertices}), we can estimate the reheating
temperature as $T_{\rm{RH}} \approx 1.3\times10^{-2} g_*^{-1/4} \left( N_s
\mu^3 / \Mpl \right)^{1/2}$, where $N_s$ is the number of minimally
coupled scalar fields into which $\phi$ decays (it decays most strongly
into these fields). Then the scale factor (normalized to $a=1$ today) 
is 
\begin{equation}
a_{\rm{end}} = 7.5 \times 10^{-32} \left(\frac{\mu}{\Mpl}\right)^{-1/6}
g_*^{-1/12} N_s^{1/6}
\end{equation}
at the end of inflation.
Integrating the slow roll equations of motion,
$\phi' = - V'(\phi)/3\tilde H$, and assuming 
$\lambda \ll 1$, 
the number of e-foldings of inflation for a mode $k$ is
\begin{equation}
\label{eq:efold}
N_k \approx  \frac{3\, \mbox{arctanh}(\sqrt{\lambda} q )}{2\sqrt{\lambda}}
- \frac{3}{4} \ln\left( 1+2 q_k \right) 
+ N_0(\lambda).
\end{equation}
Here $N_0$ is a small number defined such that $N_k(q_{end}) = 0$
at the end of inflation, where $q = q_{end}\approx 1/\sqrt{3}$,
and $q_k$ is related to the conformal factor 
$\chi_k = 1 + 2 q_k + 3 \lambda q_k^2$ when the mode $k$ crosses
the horizon:
\begin{equation}
\label{eq:crossing}
\tilde{H}_k \approx \mu/\sqrt{24} = k e^{N_k} /a_{\rm end}
\end{equation}
This particular mode will have a scalar fluctuation amplitude
(also referred to as $\delta_H^2$ in the literature)
\begin{equation}
\label{eq:amp}
Q_k^2 = \frac{1}{ 1200 \pi^2 \epsilon_k} \left(\frac{\mu^2}{\Mpl^2} \right)
\end{equation}
where the slow roll parameters (using the definitions
in \cite{Bassett:2005xm}) are
\begin{equation}
\epsilon_k \approx \frac{ (1- \lambda q_k^2)^2}{3 q_k^2}
\,, \quad 
\eta_k \approx - \frac{ 2(1 + \lambda q_k^2)}{3 q_k}.
\end{equation}
We then use these to find the the scalar spectral index $n_s=1-6\epsilon+2\eta$, 
the ratio of 
tensor to scalar modes $r=16\epsilon$ etc.
Using the combined WMAP+SDSS measurements \cite{Tegmark:2006az} 
$Q = 1.945\pm0.05 \times 10^{-5}$ for modes $k=0.002 / \rm{Mpc}$ 
we can use Eq.~(\ref{eq:efold}-\ref{eq:amp}) together to fix $\mu$.
For $\lambda\rightarrow0$ the result is 
\begin{eqnarray}
\mu&\approx&(3.2 \pm 0.1) \times 10^{-5} \Mpl,\\
n_s&\approx&0.964,\\
r  &\approx&0.0036,
\end{eqnarray}
which is consistent with the both the theoretical results of 
\cite{Hwang:2001pu,Mukhanov:1989rq} and recent observational constraints
\cite{Spergel:2006hy,Tegmark:2006az}.

In addition, $n_s$ is sensitive to the value of $\lambda$. The observational
constraint 
$0.937 < n_s <  0.969$ (68\% 
C.L.) from \cite{Tegmark:2006az},
translates into a strong upper bound on $\lambda$:
\begin{equation}
\lambda < 4.7 \times 10^{-4}.
\end{equation}
This is an example of the usual fine tuning that is needed for observationally allowed
inflationary potentials and is consistent with the findings
of \cite{Berkin:1990nu}.
More precisely the values of $\mu,\lambda$ appropriate for
inflation are shown in Fig \ref{fig:polycons}.

\subsection{Other constraints}


Above we explored the possibility that $\phi$ was the inflaton.
Let us now turn to the alternative possibility that $\phi$ is not the inflaton,
and compute miscellaneous constraints on the parameters $\mu$ and $\lambda$ when they are varied freely.
We will first consider the
fifth force mediated by $\phi$, then investigate how the scalar
field behaves dynamically in the early universe, where the
most interesting effect comes from considering
a period of slow-roll inflation driven by some \emph{other}
scalar field.
As noted in Section \ref{sec:equiv},
the dynamics of $\phi$ is still governed by an effective potential
(\ref{eq:effective}) which is important when there is a
component of matter whose energy-momentum tensor has nonzero 
trace.
To begin with, we ignore any effect that such a term may have on
the minimum of $V$ for these polynomial models, 
which is a good approximation if $|\tilde{T}^{\mu}_{\mu}|\ll
\mu^2 \Mpl^2$. We will see that for the first few 
constraints that we derive, this
will indeed be the case. Then we will return to the question of where
this is a bad approximation, which will naturally lead
to our discussion of slow-roll inflation.

\subsubsection{Fifth force constraints}


The minimum
of the effective potential lies at $\chi=1$, $\phi = 0$.
The curvature of this minimum is $m^{2}_{\phi} = \mu^2 /6$. Hence
we can get around solar system constraints simply by making
$\mu$ large enough so that the range of the fifth
force will be small. Clearly it must have an range smaller
than the solar system, otherwise, as was discussed above,
it will violate the bound on the PPN parameter $\gamma$. (Recall that
there is no Chameleon effect here, so $\Delta=1$ 
in Eq.~(\ref{eq:metric-JF}) and $\gamma=1/2$.)
For smaller scales, we consider searches for a fifth force
via deviations from the inverse square law.
The profile for a quadratic potential, i.e,
Eq.~(\ref{eq:yukawa}), gives a Yukawa potential between two
tests masses $m_1$ and $m_2$:
\begin{equation}
\label{eq:yukawa_potential}
V(r)= - \alpha {m_1 m_2 \over 8 \pi \Mpl^2} {e^{- m_\phi r}\over r},
\end{equation}
where $\alpha = 1/3$. For this $\alpha$-value, a fifth force
is ruled out for any Compton wavelength $m_\phi^{-1}$ ranging from
solar system scales down to $0.2 \rm{mm}$, where the lower
bound comes from the E\"{o}t-Wash experiments \cite{Hoyle:2000cv}.
This bound translates to 
\begin{equation}
\mu \gtrsim 1.0 \times 10^{-3} \rm{eV}.
\end{equation}
This implies 
$V(\phi) \sim \mu^2 \Mpl^2 \gg \rho_{\rm solar}$, a typical
solar system density, so 
for this constraint we were justified in ignoring any
effects of the density-dependent term on the
minimum of $V_{\rm{eff}}$.

\subsubsection{Nucleosynthesis constraints}


Niven this preliminary constraint from local gravity tests, let
us now consider the cosmology of $\phi$ in the EF. We may approximate
the potential around the minimum by a quadratic potential
$V_{\rm{eff}}(\phi) \approx (\mu^2/12) \phi^2$, which is 
valid for $| \phi | \lesssim \Mpl$. The interesting
behavior will come during the radiation dominated epoch,
so in Eq.~(\ref{eq:einstein-cosmo}) we take
$\tilde\rho(\tilde a) \approx \bar{\rho}_R(\tilde a) \propto \tilde a^{-4}$, 
and  we ignore the $\tilde{T}^\mu_\mu$ term in Eq.~(\ref{eq:scalar-cosmo}) 
to find the cosmological equations of motion
\begin{eqnarray}
\label{eq:einstein-new}
3 \tilde{H}^2 \Mpl^2 &=& \bar{\rho}_R(\tilde a)
+ \frac{\mu^2}{12} \phi^2 + \frac{1}{2} 
(\phi')^2, \\
\phi'' &+& 3 \tilde{H} \phi' + \frac{\mu^2}{6} \phi =0,
\label{eq:cosmo-new}
\end{eqnarray}
where the primes denote $d/d\tilde t$.
There are two interesting limiting behaviors, corresponding to $\tilde H\gg \mu$ and $\tilde H\ll \mu$, which we will now explore in turn.

For $\tilde H \gg \mu$, the friction term in
Eq.~(\ref{eq:cosmo-new}) dominates, and $\phi$ is frozen out
at some value $\phi_*$ with $d \phi/ d \tilde t = 0 $. 
The energy density of $\phi$ is
subdominant in Eq.~(\ref{eq:einstein-new}).  Therefore, in the EF we
have the usual radiation dominated expansion, and in the JF using 
Eq.~(\ref{eq:densitymap}) and Eq.~(\ref{eq:hubble}) we have the same
FRW expansion with a different effective Newton's constant $G_N^*$:
$3 H^2 = 8 \pi G_N^* \rho(a) \propto a^{-4}$, where
\begin{equation}
\label{eq:newt}
G_N^* = \frac{1}{8\pi \Mpl^2} \exp\left(-\sqrt{\frac{2}{3}} \frac{\phi_*}{\Mpl}
\right)
\end{equation}

For $ \tilde H \ll \mu$, on the other hand, assuming $\phi_* < \Mpl$,
the field $\phi$ starts to oscillate with frequency $\mu/\sqrt{6}$ and
an amplitude that redshifts as $\tilde a^{-3/2}$.
Hence in the EF, the
energy density of $\phi$ in Eq.~(\ref{eq:einstein-new}) from these
zero momentum field oscillations is
$\rho_\phi = (\mu^2/12) \phi^2 +  \phi'^2/2 \approx 
\rho^*_\phi (\tilde{a}/\tilde{a}_*)^{-3}$ 
, where
$\rho^*_\phi \approx (\mu^2/12) \phi_*^{2}$. Mapping back into the JF, 
and averaging
over a cycle of this oscillation, we obtain the Friedmann equation
\begin{equation}
3 H^2 \Mpl^2 = \rho_R(a) + \frac{3}{2} \rho^*_\phi (a/a_*)^{-3},
\end{equation}
where the additional factor
of $1/2$ comes from the averaging of the oscillations of $G_N^*$
in Eq.~(\ref{eq:newt}), as is discussed in more depth in \cite{Accetta:1990yb}.

The crossover between these two behaviors occurs when
$\tilde H$ is comparable to $\mu$, and given the laboratory tests of 
gravity above we can say that this must occur 
when the universe has at least the temperature 
$T_* \gtrsim 1 \rm{TeV}$. We were therefore justified in assuming radiation
domination in our calculation.

Let us examine further the zero momentum oscillations of 
$\phi$ that give this extra non-relativistic energy density. In the
absence of some mechanism (such as an extra period of low scale inflation
\cite{Randall:1994fr}), we expect the initial 
amplitude of oscillations to be of
the order $\Mpl$. 
This is because the potential in Figure \ref{fig:polypot}
varies on the scale of $\Mpl$ independently of the height of
$V$. Hence in the absence of any other scale, the initial amplitude must be 
around this size.  
Recall that at the onset of oscillations, $\tilde H \sim \mu$,
so the initial energy density of these oscillations is 
\begin{equation}
\rho^{*}_\phi \sim \Mpl^2 \mu^2 \sim \tilde H^2 \Mpl^2 \sim \rho_R(a_*).
\end{equation}
This energy density 
subsequently grows relative to the radiation density component,
quickly forcing the universe into a matter dominated period of expansion.
This is unacceptable if this component does not decay before the onset of
BBN, because  
then at the time of BBN the expansion would
be much faster than the normal radiation dominated expansion,
which would be inconsistent with observed primordial
abundances \cite{Copi:2003xd}.

The fact that $\phi$ interacts weakly with other particles
(the vertices in Eq.~(\ref{eq:vertices}) are suppressed by $1/\Mpl$)
so that $\phi$ decays too slowly
is exactly what is known as the cosmological \emph{moduli problem}. 
To be more
precise, we can use Eq.~(\ref{eq:vertices}) to estimate the decay
rate of zero momentum modes into other massive particles:
\begin{eqnarray}
\nonumber
\Gamma_\phi &\approx& \sum_s \left( \frac{m_s^4}{m_\phi \Mpl^2 96 \pi} 
- \frac{m_\phi m_s^2}{ 96 \pi \Mpl^2} + 
\frac{m^3_\phi}{ \Mpl^2 384 \pi} \right) \\ 
&& \hphantom{  \frac{m_s^4}{m_\phi \Mpl^2 96 \pi} 
+ \frac{m_\phi m_s^2}{ 96 \pi \Mpl^2} }
+ \sum_f \frac{ m_f^2 m_\phi}{\Mpl^2 12\pi},
\end{eqnarray}
where the sums are over minimally coupled scalar particles 
and fermions with masses $2 m_s, \, 2m_f < m_\phi$. The
requirement $\Gamma_\phi > H_{\rm{BBN}}$ translates into the constraint
$\mu \gtrsim 100 \,\rm{TeV}$ for the Standard Model. One would
expect the bound on $\mu$ to be slightly smaller if one includes
other particles that have not been detected yet with mass smaller
than $100 \rm{TeV}$.
This constraint should not be taken too seriously, however, because
the moduli problem may hypothetically be resolved by electroweak scale inflation
\cite{Knox:1992iy} or even by
a brief second period of inflation at the electroweak scale
\cite{Randall:1994fr}.

\subsubsection{Density dependent forces}


We now consider how the extra density dependent term in $V_{\rm eff}$
may effect cosmology.
In other words, when can we not neglect
the forcing term $\tilde{T}^{\mu}_{\mu}$ of Eq.~(\ref{eq:scalar-cosmo})?
After $\phi$ enters the oscillating phase when $\mu \gg
\tilde H$, the extra term has little effect
on the minimum since then it is small compared to the size of the potential
itself ($V \sim \mu^2 \Mpl^2$).  As a result, $\phi$ simply oscillates as
expected. Before the crossover, when $\phi$ is frozen, we showed that the
universe must be radiation dominated so that in particular,
as $\tilde{T}^{\mu}_\mu \ll \tilde \rho = 3 \tilde H^2 \Mpl^2$ 
during this phase, the Hubble friction will 
dominate compared to the force term of $\tilde{T}^{\mu}_\mu$
in Eq.~(\ref{eq:scalar-cosmo}), and we were justified 
in claiming that $\phi$ is frozen out.
The cosmology here does not suffer from the instability
that plagues Eq.~(\ref{eq:arch}).

There are, however, some exceptions that might
lead to interesting constraints.
First, consider a 
relativistic component $i$ of the cosmological plasma that becomes
non-relativistic and dumps its energy into the other
relativistic components. In this case,
$-\tilde{T}_\mu^\mu \sim (g_i/g_*)
\tilde{\rho}$  for a period
of about one e-folding, so $\phi$ receives a kick and
is displaced by an amount $\Delta \phi \approx (g_i / g_*) \Mpl/\sqrt{6}$
\cite{Brax:2004qh}. 
This might lead to an interesting effect such as
$\phi$ being kicked out of the basin of attraction
of $V$. The extreme case would be that $\phi$ does not  
end up oscillating around the minimum as expected when
$\tilde H \sim \mu$, 
but instead rolling down the tail of $V$, an effect
which is clearly only possible for $\lambda\neq 0$.
In principle, such kicks could
even invalidate the predictions of BBN: near the onset
of BBN, $e^{\pm}$ annihilation occurs, displacing $\phi$
and consequently changing $G_N^*$ significantly as per Eq.~(\ref{eq:newt}).
However, we have already shown that $\phi$
must be in the oscillatory phase long before the onset of BBN,
and we have argued that 
these kicks have no effect while $\phi$ is in the oscillator phase, so
in fact this effect is unlikely to have relevance for BBN.
Such kicks may effect
other important cosmological dynamics
at temperatures
higher than $T > 1 \rm{TeV}$, such as baryogenesis.  However,
the effects 
are extremely model dependent, and it is hard to say anything 
definitive at this point.

\subsubsection{Non-inflation}



Another situation when we cannot ignore the density
dependent force on $\phi$ is during inflation.
Here $\tilde{T}^\mu_\mu$
is large for many e-foldings. Remember that in this section, we are 
not considering $\phi$ as our inflaton;
instead we consider a slow roll inflationary period driven by some
other scalar field $\psi$ defined in the JF. 
We wish to examine the 
effect a modified gravity Lagrangian such as Eq.~(\ref{eq:polymodel}) has
on the inflationary scenario.  In particular,
we will be interested in situations where inflation by
the field $\psi$
does not work, being effectively sabotaged by $\phi$. We will discuss the
generality of these assumptions at the end.

Such models have been considered before in the context
of both the $\lambda = 0$ models \cite{PhysRevD.43.2510,Kofman:1985aw,
Cardenas:2003tg} and other generalized gravity
models \cite{Berkin:1991nm}. There the goal was generally
to make the inflationary predictions 
more successful, focusing on working models.

\begin{figure}[h!]
\hbox{\psfig{figure=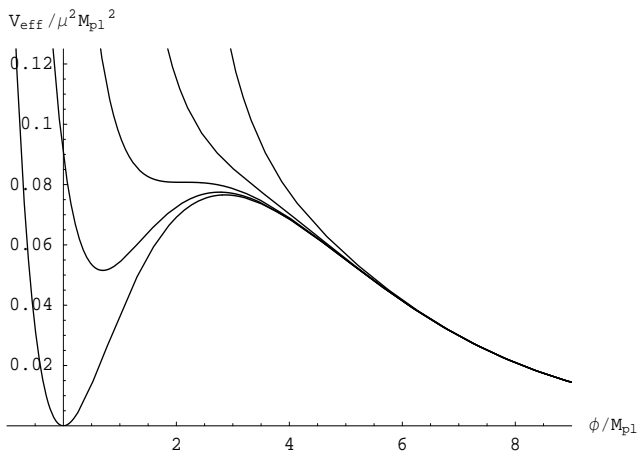,width=9cm,height=6cm}}
\caption{Effective potential for the polynomial model Eq.~(\ref{eq:polymodel}),
with various JF inflationary energy densities $u(\psi)$, here
$\lambda=0.1$.
\label{fig:polypot}}
\end{figure}

In the JF, consider a scalar field $\psi$ with a potential
$U(\psi)$. We assume that $\psi$ is slow rolling;
$d \psi/dt \approx - U'(\psi) / 3 H(t)$. 
This is the assumption that
\begin{equation}
\label{eq:slowroll}
\frac{d^2 \psi}{ d t^2} \ll U'(\psi) \, , \quad
\left(\frac{d \psi}{d t}\right)^2 \ll U(\psi)
\end{equation}
which must be checked for self-consistency 
once we have solved for $H(t)$.  We can now easily calculate
$H(t)$  by first working in the EF and mapping back to the JF.
The equations of motion in the EF 
Eq.~(\ref{eq:einstein-cosmo}-\ref{eq:scalar-cosmo}) 
become
\begin{equation}
\label{eq:infeom}
3 \tilde{H}^2 \Mpl^2 = \frac{1}{2} \phi'^2
+ V_{\rm{eff}}(\phi) \,,\quad
\phi'' + 3 \tilde{H} \phi'
 = - V'_{\rm{eff}}(\phi) 
\end{equation}
\vspace{-12pt}
\begin{equation}
\label{eq:infeffective}
V_{\rm{eff}}(\phi) = V(\phi) + U(\psi) \chi^{-2}
\end{equation}
It is interesting that a constant vacuum term in the JF does
not translate into to a constant term in the EF. See Figure \ref{fig:polypot}
for some examples of the effective potential $V_{\rm{eff}}$; we see that
for large enough $U(\psi) \gg \mu^2 \Mpl^2$, the minimum vanishes. One
finds that there is no minimum of the effective potential for
\begin{equation}
\label{eq:zeroconst}
U(\psi) > \frac{\mu^2}{18 \sqrt{3 \lambda}}.
\end{equation}
In particular, there is always a minimum for $\lambda = 0$.

The resulting behavior of the inflaton $\psi$ depends
on the size of $U(\psi)$ compared to $\mu^2 \Mpl^2$.
For small $U(\psi) \ll \mu^2 \Mpl^2$, it is clear that there is a stable minimum
around which $\phi$ will oscillate. 
In this situation, the effective potential has a minimum at
$\phi \approx 0$ with value
approximately $V_{\rm{eff}} (\phi \approx 0) \approx U(\psi)$, so after the
energy density of $\phi$ oscillations redshift away, we are left
with an exponentially expanding universe with $\chi\approx 1$, 
$3 \tilde{H}^2 \Mpl^2 \approx U(\psi)$ and $\tilde H \approx H$. 
Hence in the JF,
gravity behaves as it normally would in
general relativity: for a flat potential, the slow roll
conditions are satisfied, and inflation driven by $\psi$ 
works as it normally would. This is the
expected situation, and will happen for $\mu \approx \Mpl$.


On the other hand, we now show that when $U(\psi) \gg \mu^2 \Mpl^2$ 
and when there is no minimum of the effective potential ($\lambda
\neq 0$),
we get a contradiction to the assumption that $\psi$
was slow rolling. Hence we show that it is not possible
for $\psi$ to drive slow-roll inflation.
For large $U(\psi) \gg \mu^2 \Mpl^2$,
the potential may be approximated as 
$V_{\rm{eff}} \approx U(\psi) \chi^{-2} $.
We treat $U(\psi)$ as a constant and find
that there is an exact attractor solution to Eq.~(\ref{eq:infeom})
of the form $ \chi \sim \tilde t$ and $\tilde a \sim \tilde t^{3/4}$.
Mapping this into the EF, we find the behavior
$a \sim t^{1/2}$, i.e.,  a period of radiation dominated
expansion analogous to the $\phi$MDE of \cite{Amendola:2006kh}
More specifically, we find
\begin{equation}
3 \Mpl^2 H^2 \approx U(\psi) a^{-4}
\end{equation}
This is clearly not an inflating universe. So
the slow roll assumptions of Eq.~(\ref{eq:slowroll})
are not consistent in this case.  We therefore
conclude that it is not possible
for $\psi$ to drive slow-roll inflation.

Instead, $\psi$ dumps most of its energy $U(\psi_0)$ into radiation, and
as before, $\phi$ is left frozen at some point $\phi_*$
until $U(\psi_0) \tilde a^{-4} \sim \mu^2 \Mpl^2$. 
After this, $\phi$ can either drive an inflationary period
itself as in the original discussion of
$f(R)$ inflation, or if $\phi_*$ 
is not in the basin of attraction, it will roll down
the tail of $V$. In neither situation has $\psi$ inflated
our universe. 
From this combination of inflaton $\psi$ plus $f(R)$ gravity, we only get
satisfactory inflation 
if $(\mu,\lambda)$ lies in the region of parameter space
appropriate
for $f(R)$ inflation (the blue/grey sliver in Figure \ref{fig:polycons}) 
and if $\phi_*$ sits
at a point which allows for the required number of e-foldings.

\subsubsection{Gravitational wave constraints}


It is well-known that inflation produces horizon-scale gravitational waves of amplitude 
$Q_t\sim H/\Mpl$, so that the energy scale of inflation can be bounded from above by the
current observational upper limit 
$Q_t\simlt 0.6\times 10^{-5}$ \cite{Spergel:2006hy,Tegmark:2006az} and perhaps 
measured by a detection of the gravitational wave signal with future CMB experiments \cite{Bock:2006yf}. 
However, because of the EF-JF duality, one needs to carefully define what one means by 
``the energy scale of inflation''. 
The bound from the above argument simply
precludes inflatons with a given energy density
$U(\psi)$ in the JF, but 
$U(\psi)$ is merely a parameter which
does not necessarily set the energy scale of inflation.
In addition, we cannot derive a constraint 
for $\lambda = 0$, because
then there is always a minimum in the effective potential
and it is always possible for $\psi$ to slow roll
(this situation is described in 
greater depth in \cite{Kofman:1985aw,PhysRevD.43.2510}).

To make these ideas more concrete and resolve both
of these ambiguities, we will operationally define the energy scale of inflation 
to be the one that makes the standard GR formula for the gravitational wave amplitude valid.
It is clear that the amplitude of gravity waves
should be calculated in the EF where
the metric has a canonical action. The result is then
passed trivially into the JF after inflation and when $\phi = 0$.
The Hubble scale $\tilde H$
then sets the size of the fluctuations,
but it is a complicated
model dependent calculation to find exactly when the relevant
fluctuations are generated. However, there is a limit
to the size of $\tilde H$ for which the EF is approximately
inflating, and so gravity waves are being generated. 
Following the discussion above of non-working
inflatons, we demand that $\phi$ must
be slow rolling down the effective potential
$V_{\rm eff}$ defined in Eq.~(\ref{eq:infeffective}),
for both frames to be inflating.
In this situation, both scalar and gravity modes are being generated.

The procedure is thus to find the maximum value of $\tilde H$ 
(that is, the maximum
value of $V_{\rm{eff}}$) such that $\phi$ is slow rolling.
We then maximize this $\tilde H$ with respect to the parameter 
$U(\psi)$ to find the largest amplitude of gravity waves
that can possibly be produced. At each step in this 
procedure, we wish to be as conservative as possible; for
example, we define slow roll through the slow roll parameter constraints
$\epsilon < 1$ and $|\eta| <2$ to allows
for the possibility of power law inflation.

As an example, consider the $\lambda=0$ case. Here it is possible
to show that for $\phi$ to be slow rolling, it must satisfy
\begin{equation}
\phi > \phi_{\rm sr}\equiv \sqrt{\frac{3}{2}} \Mpl \ln \left(
\frac{2}{3} + \sqrt{\frac{7}{9} + \frac{8 U(\psi)}{3 \Mpl^2 \mu^2}} 
\right),
\end{equation}
where $\phi_{\rm sr}$ always lies to the left of the minimum
of $V_{\rm eff}$.
The maximum Hubble scale in the EF for
a given $U(\psi)$ is then 
$\tilde{H}^2 < \mbox{max}\{ V_{\rm eff}(\phi_{\rm sr})/3 \Mpl^2\, ,\,\,
\mu^2/24 \}$. This is maximized for large $U(\psi)/\mu^2\Mpl^2$,
with the result that $\tilde{H}^2< \mu^2/6$. This translates into a constraint
on the maximum gravitational wave 
amplitude
\begin{equation}
Q_t^{\rm{MAX}} \approx 0.04 \frac{\mu}{\Mpl} \,
\Longrightarrow \, r^{\rm{MAX}} \approx 5 \times 10^{6} \frac{\mu^2}{\Mpl^2}
\end{equation}
Given a measurement of 
the tensor to scalar ratio $r$, this places a limit on $\mu$:
\begin{equation}
\mu \gtrsim  3 \times 10^{-4} r^{1/2} \Mpl
\end{equation}
Numerically, we find similar
results for non-zero $\lambda$. We plot examples of this 
constraint in Figure~\ref{fig:polycons}, combined with the
already discussed working $f(R)$ inflationary models. Note that
for a given $r$, it is important that this constraint lies 
below the corresponding working inflationary model with the
same $r$; fortunately, it does.

If gravitational waves are not detected, 
then this argument gives no lower bound on $\mu$.  In particular, it is possible
that inflation occurred at the electroweak scale, in which case 
the constraint $\mu\gtrsim 2\times 10^{-3} \rm{eV}$ is
the best we can do.

Note that we completely ignore the production of scalar fluctuation modes for
this argument. This is because the scalar modes are much more difficult
to calculate, since there are two scalar fields in the mix, $\psi$
and $\phi$, which are canonically defined in different frames. But the
scalar modes are also model dependent and one should generally be able to
fine tune $U(\psi)$ to give the correct amplitude and spectral
index without affecting the above argument. This more complicated
problem was considered for chaotic inflation with
$R^2$ gravity in \cite{Cardenas:2003tg}.


This constraint applies only to slow-roll inflation models.
There are
classes of fast-roll inflation, but these models
have problems of their own and generally fail to reproduce
the required scale invariance (see \cite{Linde:2001ae} for 
a review).

Finally, let us discuss some inflaton models that might
circumvent this constraint. It is possible to add
an inflaton in the EF.  However, this theory is then not conformally
equivalent to an $f(R)$ theory: the two scalar fields $\tilde{\psi}$ 
and $\phi$ get mixed up. Hence it is not in the class of models we set
out to constrain. Another possibility is to add
an inflaton which is conformally coupled to gravity and with
a $V \propto \psi^4$ potential.  This does not change from
frame to frame and so inflation might be expected to work. 
However, it was shown by \cite{Komatsu:1997hv} that non-minimally
coupled scalar fields cannot drive inflation. 

In any case, if gravitational waves are
found, then this constraint must be thought about seriously when
using such $f(R)$ models in other astrophysical or local
gravity situations.

\section{Conclusions}
\label{sec:concl}

We have have searched for viable $f(R)$ theories 
using the wealth of knowledge on scalar scalar tensor theories
to which $f(R)$ theories are equivalent. 
We studied two classes of models:
the $f(R)$ Chameleon and massive $f(R)$ theories, which may well be the only classes of models that can be
made consistent with local gravity observations.

The $f(R)$ Chameleon that was studied is a special kind of scalar field which
hides itself from solar system tests of gravity
using non-linear effects associated with the 
all-important density dependent effective potential. 
It was shown that despite this
Chameleon behavior, solar system tests still preclude the possibility
of observably dynamical DE; the best we could do was $|w_X - 1| 
\lesssim 0.3 \times 10^{-6}$ for the effective DE equation of
state parameter $w_X$ relevant for the dynamics of the expansion.
There are of course interesting effects of the 
Chameleon both
for local gravity \cite{Khoury:2003aq} and on cosmological density perturbations
\cite{Brax:2005ew},
and these may be worth future studies 
in the context of $f(R)$ theories. 

The massive theories were found to be more relevant for very high
energy cosmology, so the conclusions were more speculative. First,
the scalar field may be the inflaton, in which case we found the required
polynomial $f(R)$ to be quite fine tuned as is usual for inflationary
potentials. 
If the scalar field was not the inflaton, then we saw that possible instabilities could spoil both inflation
and Big Bang nucleosynthesis, giving interesting constraints on the shape of $f(R)$. 
If primordial gravitational waves are detected using the CMB, then the 
most naive models of inflation have serious problems unless the
mass of the $f(R)$-scalar is very large; a measured 
scalar to tensor ratio of $r=0.05$ requires $\mu\gtrsim 7 \times 10^{-5} \Mpl$.
If gravity waves are not found, then the best we can say comes from the 
E\"{o}t-Wash laboratory experiments constraining the extent of a 
5th force; $\mu \gtrsim 2 \times 10^{-3} \rm{eV}$.


General relativity adorned with nothing but a cosmological constant, 
i.e., $f(R)=R-2\Lambda$, is a remarkable successful theory.
As we have discussed, a host of observational data probing scales from $10^{-2}$m to $10^{26}$m not only 
agree beautifully with GR, but also place sharp constraints on the parametrized departures from 
GR that we have explored.
In particular, both viable classes of $f(R)$ theories that we studied
were found to have no relevance for dynamic dark energy that is 
observationally distinguishable from ``vanilla'' dark energy, i.e., a cosmological constant. 
Since we have no good reason to believe that there are additional viable classes of $f(R)$-theories,
it appears likely that no viable $f(R)$ theories can produce the sort of interesting non-vanilla dark energy
that many observers are hoping to find.

The authors would like to thank 
Serkan Cabi, Alan Guth, Robert Wagoner and Matias Zaldarriaga
for helpful discussion.  
This work was supported by NASA grant NNG06GC55G,
NSF grants AST-0134999 and 0607597, the Kavli Foundation, the David
and Lucile
Packard Foundation, the Research Corporation
and the U.S. Department of Energy 
(D.O.E.) under cooperative research agreement DE-FC02-94ER40818.
EFB is supported by NSF Grant AST-05-07395.

\bibliography{f-of-r2.bib}

\end{document}